\documentclass[]{youtu} % For LaTeX2e

\usepackage{mathpazo}
\usepackage{graphicx}
\usepackage[numbers]{natbib}
\usepackage{hyperref}
\usepackage{url}
\usepackage{enumitem}
\usepackage{multirow}
\usepackage{booktabs}
\usepackage{multirow}
\usepackage{array}
\usepackage{longtable}
\usepackage{adjustbox}
\usepackage{tcolorbox}  % 添加这一行
\usepackage{makecell}  % 用于表格单元格内换行

\title{CUARewardBench: A Benchmark for Evaluating Reward Models on Computer-using Agent}
\author{Youtu-Agent Team$^*$}
\date{October 19, 2025}
\correspondence{haojialin@tencent.com}
\begin{document}

\abstract{
Computer-using agents (CUAs) enable task completion 
through natural interaction with operating systems and software interfaces. 
While script-based verifiers are widely adopted for evaluation, they suffer from limited scalability and inability to provide step-wise assessment. 
Reward models offer promising alternatives, but their effectiveness on CUA evaluation remains largely underexplored.
To address this gap, we present \textbf{CUARewardBench}, comprising four key contributions:
\textbf{(1) First-ever Comprehensive CUA Reward Benchmark:} 
We introduce the first benchmark 
for evaluating both outcome reward models (ORM) 
and process reward models (PRM) 
on CUA tasks, 
enabling systematic assessment across trajectory-level 
and step-level evaluation.
\textbf{(2) Diverse, Practical and Reliable Dataset:}
CUARewardBench encompasses trajectories from 10 software categories 
and 7 agent architectures with varying performance levels (25.9\%-50.8\% success rates).
All trajectories are expertly annotated through carefully designed protocols, 
% for trajectory selection, key step identification, and annotation standards,
with rigorous quality control to ensure reliability and practical applicability. 
\textbf{(3) Comprehensive Analysis and Insights:} 
Through extensive experiments 
across 7 vision-language models and 3 prompt templates, 
we reveal critical limitations of current CUA RMs, 
including insufficient visual reasoning capabilities, knowledge deficiencies, 
and the superiority of general VLMs over specialized CUA models 
for reward evaluation.
\textbf{(4) Unanimous Prompt Ensemble (UPE):}
Based on the insights from our comprehensive analysis,
we propose UPE, 
a novel ensemble method 
that significantly enhances reward model reliability 
through strict unanimous voting and strategic prompt-template configurations.
UPE achieves 89.8\% precision and 93.3\% NPV for ORM, 
and 81.7\% precision and 85.1\% NPV for PRM,
substantially outperforming single VLMs and traditional ensemble approaches.
In a short, 
this work introduces \textbf{both a comprehensive benchmark 
and a novel ensemble method} that substantially enhances CUA reward model reliability.
}
\maketitle

\renewcommand{\thefootnote}{*}
% \footnotemark % creates the * on the page where you want it  
\footnotetext{Full author list in contributions.}
\renewcommand{\thefootnote}{\arabic{footnote}}

\vspace{-.1em}

% \dateandcorrespondence

\section{Introduction}
\label{intro}

%介绍CUA
Computer-using agents (CUAs) represent a significant advancement in artificial intelligence, 
enabling large language models to interact directly with operating systems and software interfaces
to accomplish complex tasks autonomously. 
Recent advances from contemporary agents including OpenAI's Operator~\cite{openai2025operator} 
and UITARS-2~\cite{uitars2} 
have demonstrated promising performance across diverse desktop scenarios.

%介绍OSWorld如何用脚本评估
%强调CUA reward model需求：训练数据离线筛选、online rl训练需要outcome reward和step reward
%但是我们并不清楚现有的RM能达到什么水平，并且有些重要工作并没有开源使用RM的prompt和RM，亟待社区一起探索CUA RM
Evaluating CUA performance presents unique challenges that extend beyond traditional language model assessment. 
While OSWorld benchmark~\cite{osworld} initially employed manually predefined scripts for trajectory verification, 
this approach incurs prohibitively high costs of manual annotation and 
% proves 
is proved inadequate for scaling-up.
% large-scale applications. 
%这里需要一句衔接，说明VLM凭借其泛化能力可以低成本地对CUA轨迹的评估。
Consequently, VLM-based reward models (RMs) have emerged as a cost-effective alternative for trajectory evaluation.
The growing demand for CUA reward models stems from two critical needs: 
trajectory filtering to identify successful executions for off-line expert imitation as cold-start~\cite{guiowl,sea,seagent,opencua}, 
and providing reward signals for online agent reinforcement learning (RL)~\cite{seagent,uitars2}.
These models must evaluate both outcome success (whether the agent accomplished the task)
and stepwise correctness (whether individual actions contribute
% meaningfully
to the goal).
For a comprehensive review of related work, see Section~\ref{sec:related_work}. 
However, the effectiveness of existing reward models for computer-using agents remains largely unverified. 
While several recent works have proposed VLM-based reward models for CUA evaluation~\cite{guiowl,sea,uitars2}, 
many lack open-source implementations or detailed methodological descriptions, 
hindering systematic assessment and community-wide exploration of the capabilities of CUA RMs. 
This gap underscores the urgent need for standardized benchmarks 
that can rigorously evaluate and advance computer-using agent reward models.

\begin{figure}[t]
    \centering
    \includegraphics[width=\textwidth]{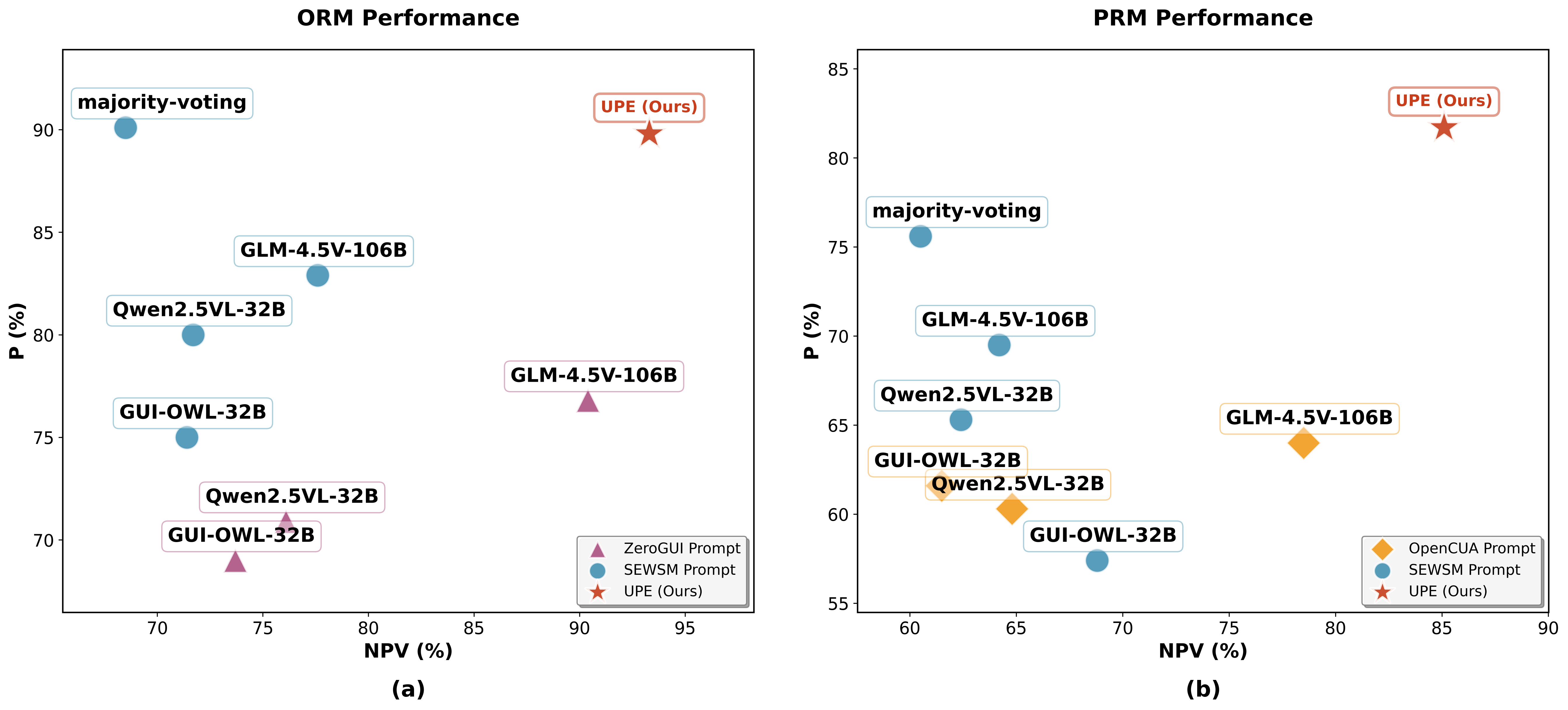}
    \vspace{-0.7cm}
    \caption{
    \textbf{UPE achieves superior reward model reliability.}
    Performance comparison on ORM and PRM tasks shows that 
    our proposed UPE (red star) simultaneously achieves high precision and NPV, 
    significantly outperforming single VLMs with different prompts 
    and traditional ensemble methods (majority-voting). 
    The upper-right positioning demonstrates UPE's effectiveness 
    in balancing positive and negative prediction accuracy. 
    Details of UPE are discussed in Section~\ref{sec:ensemble_methods}.
    }
    \label{fig:performance_comparison}
\end{figure}

\begin{figure}[t]
    \centering
    \includegraphics[width=\textwidth]{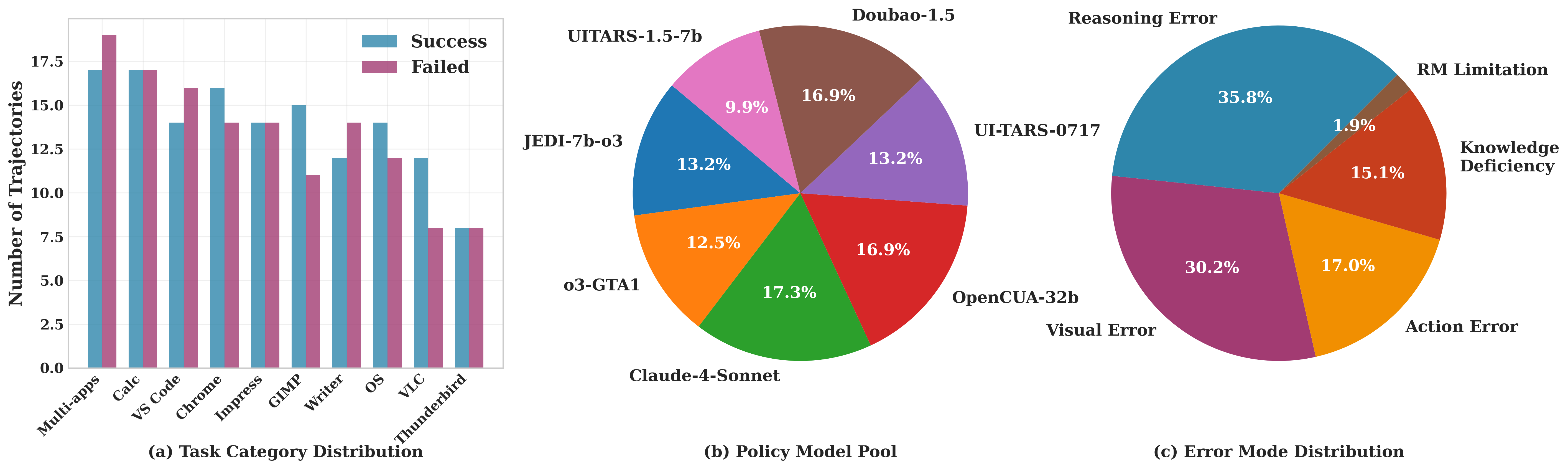}
    \caption{CUARewardBench dataset characteristics and selected experimental findings: (a) Task distribution of trajectory annotations across 10 software categories (Section~\ref{sec:annotation}), (b) policy model diversity in our benchmark (Section~\ref{sec:traj_collection}), and (c) key error modes identified in our evaluation experiments (Section~\ref{sec:error_analysis}).} 
    \label{fig:intro}
\end{figure}

To address these challenges, 
we present a systematic investigation into computer-using agent reward models 
through both benchmark construction and comprehensive evaluation. 
% 【第一段：先说benchmark构建】
Our investigation comprises three complementary components.
\textbf{First, we establish a rigorous evaluation framework 
by constructing CUARewardBench}, 
a benchmark specifically tailored to the unique requirements of CUA reward modeling 
(Figure~\ref{fig:intro}).
The benchmark design prioritizes three key aspects: 
\textit{ecological validity} through diverse desktop software interactions 
that reflect real-world agent deployment scenarios, 
\textit{comprehensive coverage} by incorporating trajectories 
from agents with varying architectural paradigms and capability levels, 
and \textit{multi-granularity assessment} enabling evaluation 
of both trajectory-level outcomes and step-level correctness. 
% 【第二段：说明实证分析】
\textbf{Second, we conduct systematic empirical analysis} 
to identify the critical factors determining reward model effectiveness, 
characterize their failure modes through detailed error analysis, 
and explore ensemble strategies that can enhance model reliability 
by leveraging complementary strengths of different approaches. 
These analyses reveal fundamental limitations in current approaches 
and provide insights into the design space for more reliable reward models.
% 【第三段：基于前两步提出UPE】
\textbf{Third, building upon these findings, 
we propose Unanimous Prompt Ensemble (UPE)}, 
a novel ensemble method that significantly enhances reward model reliability 
through strategic unanimous voting and diversified prompt-template configurations 
(Figure~\ref{fig:performance_comparison}).
As demonstrated in our experiments, 
UPE achieves substantial improvements over both single VLMs 
and traditional ensemble approaches, 
addressing the critical limitations identified through our empirical analysis.

% 【第三段：总结整体价值】
This integrated approach—from benchmark construction 
to empirical analysis to method development—enables us to provide 
both a standardized evaluation testbed for the community 
and an immediately deployable solution for enhancing CUA reward model reliability. 
The main contributions of this paper are summarized as follows:

%贡献总结：
% 1. the first comprehensive benchmark for computer-using agent ORM和PRM，
% 包含了272个轨迹成功标注和346个步骤正确标注。

% 2. 强调CUARewardBench的优点：
% 1）多样性：任务覆盖面广，并且轨迹来自各种不同的policy model Section \ref{sec:traj_collection} 
% 2）实用性&挑战性：精心设计的轨迹、step选择、标注协议，Section \ref{sec:annotation}
% 3）可靠性：大量的人工投入核查标注；

% 3. 测评对比，定量分析和定性分析（Sec X），给出了insights

\begin{itemize}[leftmargin=*]
    \item \textbf{First-ever Comprehensive CUA Reward Benchmark:} 
    We propose CUARewardBench, 
    the first comprehensive benchmark specifically designed for evaluating 
    both ORM and PRM 
    on CUA trajectories. 
    The benchmark comprises 272 trajectory success annotations 
    and 346 step correctness annotations (Table \ref{tab:annotation_distribution}),
    providing a rigorous testbed for systematic assessment 
    of reward model capabilities across both trajectory-level
    and step-level verification.

    \item \textbf{Diverse, Practical and Reliable Dataset:} 
    CUARewardBench exhibits three key strengths that establish it as a rigorous evaluation framework. 
    First, \textit{diversity}: comprehensive task coverage across 10 software categories 
    and trajectories from 7 distinct policy models with varying capabilities 
    (Section~\ref{sec:traj_collection}). 
    Second, \textit{practicality and challenge}: carefully designed protocols 
    for trajectory selection, key step identification, and annotation standards 
    that capture realistic failure modes and critical decision points. 
    Third, \textit{reliability}: extensive human validation and multi-stage quality control 
    to ensure annotation consistency and practical applicability (Section~\ref{sec:annotation}). 

    \item \textbf{Comprehensive Analysis and Insights:} 
    Through extensive experiments across 7 state-of-the-art vision-language models and 3 prompt templates, 
    we reveal that verification asymmetry challenges are significantly weakened in CUA tasks, with visual reasoning capability emerging as the overwhelmingly critical factor that dominates specialized training approaches (Section \ref{sec:reward_performance}). 
    Error analysis of 53 failure cases identifies reasoning errors (35.8\%) and visual understanding errors (30.2\%) as primary failure modes, 
    providing actionable insights for future development (Section \ref{sec:error_analysis}).

    \item \textbf{Unanimous Prompt Ensemble (UPE):}
    We propose UPE, a novel ensemble method that significantly enhances reward model reliability for CUA tasks 
    through strict unanimous voting and strategic prompt-template configurations (Section~\ref{sec:ensemble_methods}).
    As demonstrated in Figure~\ref{fig:performance_comparison},
    UPE achieves 89.8\% precision and 93.3\% NPV for ORM, 
    and 81.7\% precision and 85.1\% NPV for PRM,
    substantially outperforming single VLMs with different prompts and traditional ensemble methods such as majority voting.
    This contribution provides a practical and immediately deployable solution 
    for improving the reliability of reward-based CUA training pipelines.
\end{itemize}

\section{CUARewardBench}
\label{sec:cuarewardbench}
This section presents the construction and characteristics of CUARewardBench, 
our comprehensive benchmark for evaluating reward models in computer-using agent tasks. 
We begin by formalizing the problem setting and defining key concepts 
in Section~\ref{sec:problem_formulation}. 
Section~\ref{sec:traj_collection} details our trajectory collection methodology, 
including task selection, policy model diversity, 
and data curation protocols. 
Section~\ref{sec:annotation} describes our rigorous annotation process, 
quality control measures, and the resulting dataset statistics. 
Figure~\ref{fig:intro} provides an overview of the dataset characteristics 
and key findings from our evaluation experiments.

\subsection{Problem Formulation}
\label{sec:problem_formulation}
\textbf{Trajectory Definition.} 
Let $q$ denote the instruction given by user, 
$o_i$ denote a system state observation at step $i$, 
and $a_i$ represent an executable action in an operating environment $\mathcal{E}$ such that $o_{i+1} = \mathcal{E}(o_i, a_i)$. 
A computer-using agent trajectory is defined as the sequence:
\begin{equation}    
\mathcal{T} = \{q, o_1, (r_1, a_1), o_2, (r_2, a_2), \dots, (r_{n-1}, a_{n-1}), o_n \}
\end{equation}

where $r_i$ is the agent's reasoning for action $a_i$, and $o_n$ is the terminal state. 
Each state observation $o_i$ contains a \textbf{single RGB screenshot} of the current interface, consistent with current reward model practices that use visual inputs exclusively.

\textbf{Reward Model Formulation.}
Given trajectory $\mathcal{T}$, a reward model $\mathcal{R}$ predicts:
\begin{equation}    
\hat{\mathcal{A}} = \mathcal{R}(\mathcal{T}) = (\hat{s}, \{\hat{c}_1, \dots, \hat{c}_{n-1}\})
\end{equation}
where $\hat{s} \in \{0,1\}$ indicates 
trajectory success ($1 = \text{success}$, $0 = \text{failure}$), 
and $\hat{c}_i \in \{0,1\}$ denotes 
step correctness ($1 = \text{correctness}$, $0 = \text{non-correctness}$). 
In recent researches \cite{zerogui,seagent,sea,guiowl,uitars2}, 
$\mathcal{R}$ is exclusively implemented using VLMs as the core evaluation engine.

The following sections detail the construction of CUARewardBench: Section~\ref{sec:traj_collection} describes the curation process for trajectories $\mathcal{T}$, while Section~\ref{sec:annotation} presents the ground truth annotation methodology for $\hat{s}$ and $\hat{c}_i$.

\subsection{Trajectory Collection.}
\label{sec:traj_collection}
\textbf{Tasks and Environments.}
%就说用了osworld，选了10个task categories。
%先说为啥用osworld，就说这个benchmark用得广泛啥啥的，然后开始介绍osworld
%osworld包括10个常用的软件，task categories（vs_code, gimp, libreoffice_writer, chrome, multi_app
%libreoffice_impress, libreoffice_calc, vlc, thunderbird, os)
%对这10个task categories，我们那些infeasible的task，只在feasible task中选择轨迹。
We build CUARewardBench upon OSWorld~\cite{osworld}, 
a widely-adopted benchmark that provides comprehensive evaluation environments 
for computer-using agents across diverse desktop applications. 
OSWorld is particularly well-suited for our purposes 
due to its widespread adoption in the research community 
and its comprehensive coverage of realistic computer interaction scenarios. 
The benchmark encompasses 10 common software applications 
across different task categories, 
including Chrome, Thunderbird, LibreOffice Writer, LibreOffice Calc, LibreOffice Impress, 
VS Code, GIMP, VLC, and OS operations,
providing a rich ecosystem for evaluating computer-using agent capabilities.
CUARewardBench cover all 10 task categories to provide comprehensive evaluation 
across the complete spectrum of computer-use scenarios.
To ensure benchmark quality and maintain evaluation reliability, 
we systematically exclude tasks marked as infeasible 
in the original OSWorld dataset 
to avoid introducing evaluation noise from inherently unsolvable scenarios.

\textbf{Policy Model Pool.}
%过渡一句，和上一段自然衔接
%为了保证生产的轨迹的多样性，选择了7个agent，共计10个configuration（有的agent采用了不同的max step limit）
To ensure trajectory diversity and comprehensive coverage of agent capabilities, 
we employ 7 distinct CUA models with 10 different configurations 
(some agents vary in maximum step limits). 
%As show in Table 1, 我们主要从两方面来选择agent：
% 1）不同的CUA架构（single model到agentic framework）
% 2）不同的performance(25.9\% to 50.8\% success rates on OSWorld
% 一方面，两个维度的考虑不仅使得轨迹的多样性得到保证，capture a wide spectrum of decision-making patterns and failure modes, 
% providing a robust foundation for evaluating reward model generalization
% 另一方面，agent池足够大也同时也使得成功轨迹对task的覆盖率得到提升，（Oracle SR达到72.3），
% 这为我们后续挑选轨迹提供了充足的数据基础。
As shown in Table~\ref{tab:agent_performance}, our agent selection follows two key principles: 
1) \textit{Architectural Diversity}: We include both single-model approaches and agentic frameworks to capture different decision-making paradigms.
2) \textit{Performance Spectrum}: Our selected agents span success rates from 25.9\% to 50.8\% on OSWorld, ensuring comprehensive coverage of capability levels.

This dual consideration ensures trajectory diversity by capturing a wide spectrum of decision-making patterns and failure modes, providing a robust foundation for evaluating reward model generalization. Additionally, our diverse agent pool enhances task coverage for successful trajectories, achieving an oracle success rate of 72.3\% and establishing a solid data foundation for subsequent trajectory curation. 

% ============================================================
%                      model_setting 分布统计                     
% ------------------------------------------------------------
% claude-4-sonnet-20250514                    : 47
% opencua_agent-opencua_32b                   : 46
% doubao-1-5-thinking-vision-pro-250428-15step: 46
% jedi-7b-o3                                  : 36
% UI-TARS-0717-15step                         : 36
% o3_gta1_50steps                             : 34
% uitars15-7b-15step                          : 27
% ------------------------------------------------------------
% TOTAL                                       : 272
% ============================================================

\begin{table}[t]
    \centering
    \footnotesize  % 改为更小的字体
    \setlength{\tabcolsep}{6pt}  % 减小列间距
    \begin{tabular}{lcccc}
    \toprule
    Agent Model & Arch & SR (15-S) & SR (50-S) & Traj \\
    \midrule
    JEDI-7b-o3 \cite{jedi} & Framework & 42.4 & 50.8 & 36 \\
    o3-GTA1 \cite{gta1} & Framework & - & 48.8 & 34 \\
    Claude-4-Sonnet \cite{ClaudeComputerUse} & Single & 31.3 & 44.0 & 47 \\
    OpenCUA-32b \cite{opencua} & Single & 28.3 & 33.9 & 46 \\
    UI-TARS-0717 \cite{uitars} & Single & 31.9 & - & 36 \\
    Doubao-1-5-Thinking \cite{seed15} & Single & 28.3 & - & 46 \\
    UITARS-1.5-7b \cite{uitars} & Single & 25.9 & - & 27 \\
    \midrule
    \textbf{Oracle SR / Total Num} & - & \multicolumn{2}{c}{\textbf{72.3}} & \textbf{272} \\ 
    \bottomrule
    \end{tabular}
    \caption{Policy model pool of for dataset curation: agent architecture, success rates on OSWorld across different step limits, and  trajectory collection counts.}
    \label{tab:agent_performance}
\end{table}

\textbf{Trajectory Selection Criteria.} 
%过渡一句，和上一段自然衔接，
%然后提到osworld-verified已经提供了现成的轨迹，我就按照前文所确定的5个任务类型、7个agent，
%在这些生产好的轨迹\footnote{huggingface url}中进行挑选，
%挑选轨迹的标准，总结为以下几个方面：
%Task Categories Balance：保持任务类型分布平衡，这样方便我们观察reward model在不同任务类型之间的表现差异
%Success-failure balance：保持成功轨迹和失败轨迹的平衡，这样方便观察reward model对成功轨迹和失败轨迹的判别能力
%Difficulty控制：为了排除难度太低和难度太高的轨迹，对于每条轨迹，如果成功的agent configuration数量为0或者大于等于8，我们则不予考虑。
%Task Progress：排除任务进度低的轨迹，因为进度低的轨迹，一方面判定traj-success相对简单，另一方面，缺乏有效的指令-动作关联，反倒是造成action correctness的模糊。
%Step count：我们选择小于25步以内的轨迹。这是因为，25步足够大多数任务完成，出于标注成本考虑，我们选择25步以内
%Expertise-Bound Exclusion：要知道即便是Human在osworld上的成功率也只有72.36%\cite{osworld}，
%因此对于少数超出标注者能力的任务和或是正确性难以判别的轨迹，我们不纳入考虑
Building upon the established agent configurations, 
we leverage pre-collected trajectories \footnote{\url{https://huggingface.co/datasets/xlangai/ubuntu_osworld_verified_trajs}}
from OSWorld-verified \cite{osworld_verified} 
across all 10 task categories and 7 agent models. 
Our trajectory curation follows systematic criteria 
designed to ensure benchmark quality and evaluation comprehensiveness:

\begin{itemize}[leftmargin=*]
    \item \textit{Task Category Balance:} We maintain balanced distribution across software categories to enable comparative analysis of reward model performance.
    \item \textit{Success-Failure Balance:} We ensure proportional coverage of successful and failed trajectories to evaluate reward models' discriminative capabilities across both outcome types. 
    \item \textit{Difficulty Control:} We exclude trajectories 
    where no agent succeeds (too difficult) 
    or where 8+ agent configurations succeed (too easy), 
    ensuring moderate difficulty levels 
    that provide meaningful evaluation challenges.     
    % \item \textit{Task Progress Filtering:} We exclude trajectories with minimal task progress, as low-progress trajectories present overly simplistic success determination while lacking meaningful instruction-action associations that would otherwise create ambiguity in action correctness evaluation.
    \item \textit{Step Count Constraint:} We select trajectories 
    containing fewer than 25 steps, 
    as this threshold accommodates most task completions 
    while maintaining manageable annotation costs.
    % \item \textit{Expertise-Bound Exclusion:} Given that human performance on OSWorld achieves only 72.36\% success rate~\cite{osworld}, we exclude tasks that exceed annotator capabilities or present ambiguous correctness determination to maintain annotation reliability.
\end{itemize}

\subsection{Annotation}
\label{sec:annotation}
\textbf{Trajectory Success.}
%第一段是介绍轨迹标注的标准
%尽管OSWorld本身提供了基于脚本的轨迹成功评估，但是为了保证标注的可靠性，
%我们还是用人工方式来埃个评估轨迹的成功与否。
%标注人员衡量Trajectory Success的标准概括为两条：
%1）Instruction一致性：agent是否将computer的最终状态转移至指令所要求的状态
%2）Harmful Side effects：agent是否对computer的最终状态产生了指令没有要求的影响，并且消除影响需要额外的action 执行。
Although OSWorld provides script-based trajectory success evaluation, 
we employ human annotation to ensure annotation reliability and accuracy. 
Annotators evaluate trajectory success based on two primary criteria:
\begin{itemize}[leftmargin=*]
    \item \textit{Instruction Consistency}: Whether the agent successfully transitions the computer's final state to match the requirements specified in the instruction.
    
    \item \textit{Harmful Side Effects}: Whether the agent 
    causes unintended changes to the computer's final state 
    that are not required by the instruction 
    and would require additional corrective actions to resolve.
    Note that redundant but harmless operations (e.g., extra clicks on desktop) are not considered violations of this criterion. 
\end{itemize}

% ================================================================================
% CuaRewardBench 数据统计结果
% ================================================================================
% Task Type            Traj Annos         Action Annos             
%                      Success  Failure  Good         Bad         
% --------------------------------------------------------------------------------
% Multi-Apps           17       19       23           22          
% Calc                 17       17       20           17          
% VSCode               14       16       23           19          
% Chrome               16       14       24           24          
% Impress              14       14       20           16          
% GIMP                 15       11       20           16          
% Writer               12       14       15           14          
% OS                   14       12       13           15          
% VLC                  12       8        16           10          
% Thunderbird          8        8        8            11          
% --------------------------------------------------------------------------------
% Total                139      133      182          164         

\begin{table}[t]
    \centering
    \footnotesize  % 改为更小的字体
    \setlength{\tabcolsep}{3.8pt}  % 减小列间距
    \begin{tabular}{lcccc}
    \toprule
    Task Category & Success Traj. & Failed Traj. & Good Actions & Bad Actions \\
    \midrule
    Multi-apps & 17 & 19 & 23 & 22 \\
    LibreOffice Calc & 17 & 17 & 20 & 17 \\
    VS Code & 14 & 16 & 23 & 19 \\
    Chrome & 16 & 14 & 24 & 24 \\
    LibreOffice Impress & 14 & 14 & 20 & 16 \\
    GIMP & 15 & 11 & 20 & 16 \\
    LibreOffice Writer & 12 & 14 & 15 & 14 \\
    OS & 14 & 12 & 13 & 15 \\
    VLC & 12 & 8 & 16 & 10 \\
    Thunderbird & 8 & 8 & 8 & 11 \\
    \midrule
    \textbf{Total} & \textbf{139} & \textbf{133} & \textbf{182} & \textbf{164} \\
    \bottomrule
    \end{tabular}
    \caption{Annotation distribution of trajectories and actions across task categories in CUARewardBench.}
    \label{tab:annotation_distribution}
\end{table}

\textbf{Step Correctness.}
%标准就一个：该step的执行结果是否对轨迹成功产生正面影响。符号化地表述，TS为轨迹成功的状态，
%如果 p( TS | o_i+1, a_i) > p( TS | o_i)，那么action a_i是正确的。
%显然，对于错误step，p( TS | o_i+1, a_i) < p( TS | o_i)，即需要额外的action来消除a_i的负面影响。
%而对于冗余action，p( TS | o_i+1, a_i) = p( TS | o_i)，
%但是在标注过程中我们发现，冗余动作的危害远小于错误动作，并且冗余动作的判别难度高于错误动作，
%因此为了提高标注的客观性，我们的标注中尽量不考虑冗余动作，只标注准确action和错误action。
%此外，我们并不是对\ref{sec:traj_collection}中选定轨迹的每一个action都进行标注，
%我们只选取那些对轨迹成功较为关键的步骤，形式化地，
%key good action可以定义为：p( TS | o_i+1, a_i) - p( TS | o_i) 尽可能地大且 p(a_i ｜ o_i) 尽可能地小，
%直观地理解就是，该动作对任务成功的推进幅度较大且较难想到。
%key good action可以定义为：|p( TS | o_i+1, a_i) - p( TS | o_i)| 尽可能地大且 p(a_i ｜ o_i) 尽可能地大，
%直观地理解就是，该动作对任务成功的破坏幅度较大且迷惑性较强。
We evaluate step correctness based on a single criterion: whether the step's execution positively contributes to trajectory success. Formally, let $o_{\text{ts}}$ denote the trajectory success event. An action $a_i$ is considered correct if $p(o_{\text{ts}} | o_{i+1}, a_i) > p(o_{\text{ts}} | o_i)$. Conversely, for incorrect steps, $p(o_{\text{ts}} | o_{i+1}, a_i) < p(o_{\text{ts}} | o_i)$, indicating that additional actions are required to mitigate the negative effects of $a_i$. For redundant actions, $p(o_{\text{ts}} | o_{i+1}, a_i) = p(o_{\text{ts}} | o_i)$.
However, during our annotation process, we observed that redundant actions cause significantly less harm than incorrect actions, while being considerably more difficult to identify reliably. To enhance annotation objectivity, we focus primarily on distinguishing between correct and incorrect actions, largely excluding redundant actions from our evaluation.
Furthermore, we do not annotate every action in the trajectories selected from Section~\ref{sec:traj_collection}. Instead, we focus on steps that are critical to trajectory success. Formally, we define two types of key actions:
\begin{itemize}[leftmargin=*]
    \item \textit{Key Good Actions}: Actions where $p(o_{\text{ts}} | o_{i+1}, a_i) - p(o_{\text{ts}} | o_i)$ is as large as possible and $p(a_i | o_i)$ is as small as possible. Intuitively, these represent actions that significantly advance task completion while being non-obvious choices.
    
    \item \textit{Key Bad Actions}: Actions where $|p(o_{\text{ts}} | o_{i+1}, a_i) - p(o_{\text{ts}} | o_i)|$ is as large as possible and $p(a_i | o_i)$ is as large as possible. These represent actions that substantially hinder task success while appearing deceptively reasonable.
\end{itemize} 

% \textbf{Annotator Workflow.}
%为了保证标注的质量，我们将标注劳动力分为两组
%每组对各自分配到的轨迹进行标注，对错误的轨迹和错误的action要注明原因。
%而后，两组人员再交叉检查，对不能达成一致的标注进行讨论或舍弃。
% To ensure annotation quality and reliability, we organize our annotation workforce into two independent teams. Each team is assigned a distinct subset of trajectories and conducts comprehensive annotation, providing detailed justifications for failed trajectories and incorrect actions. Subsequently, both teams perform cross-validation on each other's annotations, engaging in discussion to resolve disagreements or excluding ambiguous cases where consensus cannot be reached. This dual-team approach with cross-validation ensures robust annotation quality and minimizes individual annotator bias.
% To ensure annotation quality and reliability, we organize our annotation process into two independent phases. In the first phase, trajectories are systematically annotated with comprehensive justifications provided for failed trajectories and incorrect actions. Subsequently, a second phase involves thorough cross-validation of all annotations, where disagreements are carefully reviewed and ambiguous cases that cannot be definitively resolved are excluded from the final dataset. This dual-phase approach with comprehensive cross-validation ensures robust annotation quality and minimizes potential annotation bias through systematic review processes.

\textbf{Annotation Statistics.}
%承接一下上一段，自然过渡到本段介绍标注的统计分布
%强调一下，虽然OSWorld提供了轨迹成功评估，但是我们还是用人工标注来保证标注的可靠性。
%并且对于action的标注，我们按照上一段的描述挑选了key action进行标注。
%并且我们也会在成功轨迹中标注bad action、在失败轨迹中标注good action。
%最后看图说话介绍一下成功轨迹、失败轨迹、good action、bad action的总数。
As shown in Table~\ref{tab:annotation_distribution}, our final dataset comprises 139 successful and 133 failed trajectories for trajectory-level evaluation, and 182 good and 164 bad actions for action-level assessment across all task categories.
For trajectory-level annotations, 
although OSWorld~\cite{osworld} provides script-based trajectory success evaluation, 
we employ human annotation 
to ensure annotation reliability and maintain high-quality standards. 
For action-level annotations, 
we selectively annotate key actions following the criteria 
outlined in the previous section, 
focusing on steps that are most critical for trajectory success evaluation.
Notably, we further annotate bad actions within successful trajectories 
and good actions within failed trajectories, 
recognizing that trajectory success and step correctness are orthogonal.
This enables robust evaluation of reward models across varying trajectory contexts.

\section{Reward Performance and Analysis}
\label{sec:reward_performance}
%总体起一句，就说这节先介绍评估如何的实现
%然后随后几节是从宏观分析各个因素对reward model的影响

\subsection{Implementations}
%这里要先点明，Reward Model的实现分为VLM和Prompt两方面，
%我们构建CUARewardBench，要同时测评prompt和VLM这两个维度。
Reward model implementation involves two key dimensions: VLM and prompt. 
Our CUARewardBench systematically evaluates both to provide comprehensive insights into CUA RM.

% 考虑到CUARewardBench的实际应用场景——大规模数据构造和在线强化学习训练，
% 我们主要对开源模型进行评估，这不仅是为了降低复现本文研究的实现难度，也是考虑到了在实际应用中的大规模部署的可行性。 
%我们主要测评4类模型（用\item格式,[leftmargin=*],\textit）:
%1）通用VLM：
%介绍Qwen2.5VL系列模型 \cite{qwen2_5vl}，开源模型中的leading，
%我们对系列中的7B、32B、72B进行评估

%2）visual reasoning model：
% 介绍GLM-4.5V-106B \cite{glm4_5v}，MOE激活12B，视觉推理模型，
% 不仅具备通用视觉理解能力，并且具备较强的CUA能力（OSWorld得分35.8）

%3）CUA-specialized VLM：
% CUA model被赋予了大量的CUA相关知识，直觉上来说应该具备CUA轨迹评估能力。
% 因此我们也尝试测评CUA model作为RM的效果。
% 大部分CUA model甚至无法follow指令去评估轨迹，
% 我们猜测是后训练中大量CUA数据导致了模型的灾难性遗忘
% 但是GUI-OWL\cite{guiowl}系列是个例外，based on Qwen2.5VL
% 后训练数据混合了CUA数据训练和general reasoing数据，
% 因此在获得较强CUA能力（GUI-OWL-7B模型在osworld上的得分为29.4））的同时保持较强的reasoning能力，
% 我们也测评GUI-OWL-7B和32B作为RM的效果。

%4）专用CUA reward model：SE-WSM
% SE-WSM \cite{seagent}是现有唯一开源的CUA-specialize reward model，基于Qwen2.5VL-7B，
% 经过专门的CUA reward training。
% 其训练数据是源自43 feasible chrome tasks in OSWorld，
% 经过UI-TARS and Gemini-2.5-Pro在这些任务上跑出的860条轨迹，并经过GPT-4o标注。
\textbf{VLM Selection .} 
Considering the practical application scenarios of CUARewardBench—large-scale data construction and online reinforcement learning training—we primarily evaluate open-source models. This choice not only reduces the implementation difficulty for reproducing our research but also considers the feasibility of large-scale deployment in real-world applications.
We evaluate four categories of models (7 models in total):
\begin{itemize}[leftmargin=*]
    \item \textit{General VLMs}: We assess the Qwen2.5VL series~\cite{qwen2_5vl}, which represents the leading open-source vision-language models. We evaluate three variants within this series: 7B, 32B, and 72B parameters.
    
    \item \textit{Visual Reasoning Models}: We include GLM-4.5V-106B~\cite{glm4_5v}, a mixture-of-experts model with 12B activated parameters. This model not only possesses general visual understanding capabilities but also demonstrates strong CUA performance (achieving 35.8\% on OSWorld).
    
    \item \textit{Specialized CUA Models}: CUA models are endowed with extensive CUA-related knowledge, intuitively suggesting they should possess CUA trajectory evaluation capabilities. However, most CUA models fail to follow instructions for trajectory evaluation, which we attribute to catastrophic forgetting caused by extensive CUA data during post-training. GUI-OWL~\cite{guiowl} series represents an exception, built upon Qwen2.5VL with post-training data mixing CUA training and general reasoning data. This approach enables strong CUA capabilities (GUI-OWL-7B achieves 29.4\% on OSWorld) while maintaining robust reasoning abilities. We evaluate both GUI-OWL-7B and 32B as reward models.
    
    \item \textit{Specialized CUA Reward Models}: World State Model of SE-Agent~\cite{seagent} (SE-WSM) is the only existing open-source CUA-specialized reward model, based on Qwen2.5VL-7B with dedicated CUA reward training. Its training data comprises 860 trajectories from 43 feasible Chrome tasks in OSWorld, executed by UI-TARS and Gemini-2.5-Pro, and automatic annotated by GPT-4o.
\end{itemize}

%将 prompt template开源的prompt template有这三个工作（用\item格式,[leftmargin=*],\textit）:
% 1) ORM prompt of ZeroGUI ~\cite{zerogui}. 
% ZeroGUI设计了一个先逐帧caption再整体分析的prompt template模板，
% 要求Qwen2.5VL-32B评估轨迹是否完成任务。
%我们采用ZeroGUI的prompt template作为CUARewardBench的ORM prompt template。

% 2) step reflector prompt of OpenCUA ~\cite{opencua}.
% OpenCUA在标注思维链的数据步骤之一是将claude作为reflector，
% 根据前序步骤的cot和当前截图对生成当前步骤的反思。
% 考虑到reflector和前序步骤cot的耦合的复杂性和对比实验的公平性
% 我们对reflector prompt进行了一些简化后作为CUARewardBench的PRM prompt template
% 具体的简化前后对比见附录。

% 3) prompt of SE-WSM ~\cite{seagent}.
% SE-WSM对输入的轨迹进行step-by-step的分析，然后给出多维度的评估，
% 包括轨迹正确性、冗余步骤、first error step、correct action suggestion。
% 这种同时覆盖coarse和fine-grained的评估，使其同时具备了ORM和PRM的功能。
%我们采用 SE-WSM 的prompt template同时作为CUARewardBench的ORM和PRM的prompt template。

%详细的prompt template以及模型输出的解析方式见附录
\textbf{Prompt Template.} 
Among existing works, we identify three that have open-sourced their prompt templates for CUA reward modeling: 
\begin{itemize}[leftmargin=*]
    \item \textit{ORM prompt of ZeroGUI}~\cite{zerogui}: ZeroGUI designs a frame-by-frame captioning followed by holistic analysis prompt template, requiring Qwen2.5VL-32B to evaluate whether trajectories accomplish their tasks. We adopt ZeroGUI's prompt template as CUARewardBench's ORM prompt template.
    
    \item \textit{Step reflector prompt of OpenCUA}~\cite{opencua}: OpenCUA employs Claude as a reflector in their chain-of-thought data annotation pipeline, generating reflections for current steps based on previous step reasoning and current screenshots. Considering the coupling complexity between the reflector and previous step reasoning, and to ensure fair comparative experiments, we simplify the reflector prompt as CUARewardBench's PRM prompt template. The detailed comparison before and after simplification is provided in Section \ref{sec:prompts}.
    
    \item \textit{Prompt of SE-WSM}~\cite{seagent}: SE-WSM conducts step-by-step analysis of input trajectories, providing multi-dimensional evaluations including trajectory correctness, redundant steps, first error step, and correct action suggestions. This comprehensive evaluation covering both coarse and fine-grained assessments enables it to function as both ORM and PRM. We adopt SE-WSM's prompt template for both ORM and PRM evaluation in CUARewardBench.
\end{itemize}
Detailed prompt templates and model output parsing methods are provided in Section \ref{sec:prompts} .

\begin{table}[t]
    \centering
    \footnotesize  % 改为更小的字体
    \begin{tabular}{lccc}
    \toprule
    \textbf{Metric}  & \textbf{Formula} & \textbf{Interpretation} & \textbf{Assessment Aspect} \\
    \midrule
    \multirow{2}{*}{Precision} & \multirow{2}{*}{$TP / (TP + FP)$} & proportion of predicted positive cases  & \multirow{2}{*}{reward reliability} \\
              &                & that are actually positive & \\
    \midrule
    \multirow{2}{*}{NPV} & \multirow{2}{*}{$TN / (TN + FN)$} & proportion of predicted negative cases  & \multirow{2}{*}{reward reliability} \\
     &                    & that are actually negative & \\
    \midrule
    \multirow{2}{*}{Recall} & \multirow{2}{*}{$TP / (TP + FN)$} & proportion of actual positive cases  & \multirow{2}{*}{sample efficiency} \\
                        &                 & that are correctly predicted & \\
    \midrule
    \multirow{2}{*}{Specificity} & \multirow{2}{*}{$TN / (TN + FP)$} & proportion of actual negative cases  & \multirow{2}{*}{sample efficiency} \\
                &                & that are correctly predicted & \\
    \bottomrule
    \end{tabular}
    \caption{Evaluation metrics for CUA reward models. TP = True Positive (correctly predicted positive cases), FP = False Positive (incorrectly predicted positive cases), TN = True Negative (correctly predicted negative cases), FN = False Negative (incorrectly predicted negative cases).}
    \label{tab:evaluation_metrics}
\end{table}

\textbf{Evaluation Metrics.}
Existing evaluation frameworks have adopted different metric combinations for assessing reward models. AgentRewardBench~\cite{lu2025agentrewardbench} primarily focuses on precision, while SE-WSM~\cite{seagent} employs precision and NPV (Negative Predictive Value). To comprehensively evaluate reward model performance, we consider two primary use cases: offline trajectory filtering and online reward provision in reinforcement learning. 
In the first scenario, precision reflects the proportion of trajectories correctly identified as successful by the reward model, indicating reward reliability. 
Recall measures the coverage of actual positive trajectories, representing sample efficiency. 
For the second scenario, since negative samples with low rewards also contribute to model updates in RL, we introduce NPV and Specificity metrics. 
These metrics serve as counterparts to precision and recall, reflecting the reliability of negative reward signals and sample efficiency for negative cases, respectively.

As summarized in Table~\ref{tab:evaluation_metrics}, we employ four complementary metrics to comprehensively assess reward model performance. 
Since reward reliability is more critical than sample efficiency in both offline trajectory filtering and online RL scenarios, we prioritize precision and NPV as primary evaluation metrics, with recall and specificity serving as secondary indicators in our subsequent benchmarking analysis.

% Similar to AgentRewardBench~, 
% we prioritize precision as the primary evaluation metric, 
% with recall as a secondary consideration. 
% This emphasis reflects the practical application scenarios of reward models 
% in trajectory filtering. 
% Precision (the proportion of true positives among predicted positives) 
% indicates the quality of trajectories selected by the reward model, 
% while recall (the proportion of true positives among actual positives) 
% reflects the efficiency of trajectory selection. 
% The former measures data quality in production, 
% while the latter measures production efficiency.

\begin{table*}[t]
    \centering
    \scriptsize  % 改为更小的字体
    \setlength{\tabcolsep}{3.8pt}  % 减小列间距以容纳更多列
    \begin{tabular}{lcccccc|cc|cc|cc|cc|cc}
    \toprule
    \multirow{2}{*}{Reward Model} & \multicolumn{6}{c|}{Overall} & \multicolumn{2}{c|}{vscode} & \multicolumn{2}{c|}{gimp} & \multicolumn{2}{c|}{writer} & \multicolumn{2}{c|}{chrome} & \multicolumn{2}{c}{multi\_apps} \\
    & P & NPV & R & S & OA & & P & NPV & P & NPV & P & NPV & P & NPV & P & NPV \\
    \midrule
    \multicolumn{17}{c}{\textit{zerogui}} \\
    \midrule
    Qwen2.5VL-7B & 60.8 & 65.0 & 74.8 & 49.2 & 62.1 & & 47.8 & 57.1 & 57.1 & 40.0 & 46.7 & 54.5 & 72.2 & 72.7 & 93.3 & 85.7 \\
    Qwen2.5VL-32B & 70.9 & 76.1 & 80.6 & 65.2 & 72.8 & & 75.0 & 85.7 & 68.8 & 60.0 & 53.3 & 63.6 & 70.0 & 80.0 & 85.0 & 100.0 \\
    Qwen2.5VL-72B & 70.0 & 72.1 & 75.5 & 66.2 & 71.0 & & 65.0 & 90.0 & 70.6 & 66.7 & 60.0 & 62.5 & 66.7 & 77.8 & 88.9 & 94.4 \\
    GLM-4.5V-106B & \textbf{76.8} & \textbf{90.4} & \textbf{92.8} & \textbf{70.7} & \textbf{82.0} & & \textbf{73.7} & \textbf{100.0} & \textbf{75.0} & \textbf{100.0} & \textbf{66.7} & \textbf{81.8} & 72.7 & \textbf{100.0} & 89.5 & \textbf{100.0} \\
    GUI-OWL-7B & 69.0 & 68.8 & 71.9 & 65.6 & 68.4 & & 70.6 & 83.3 & 61.9 & 60.0 & 64.3 & 75.0 & 66.7 & 77.8 & 88.2 & 88.9 \\
    GUI-OWL-32B & 69.0 & 73.7 & 78.4 & 63.2 & 71.0 & & 64.3 & 68.8 & 65.0 & 66.7 & 61.5 & 69.2 & \textbf{73.7} & 81.8 & \textbf{100.0} & \textbf{100.0} \\
    \midrule
    \multicolumn{17}{c}{\textit{sewsm}} \\
    \midrule
    Qwen2.5VL-7B & 63.1 & 57.1 & 50.4 & 69.2 & 59.6 & & 50.0 & 54.5 & 66.7 & 47.1 & 46.7 & 54.5 & 70.6 & 69.2 & 84.6 & 73.9 \\
    Qwen2.5VL-32B & 80.0 & 71.7 & 69.1 & 82.0 & 75.4 & & 84.6 & 82.4 & 75.0 & 57.1 & 63.6 & 66.7 & 75.0 & 71.4 & 87.5 & 85.0 \\
    Qwen2.5VL-72B & 78.6 & 74.5 & 74.1 & 78.9 & 76.5 & & 85.7 & \textbf{87.5} & \textbf{78.6} & 66.7 & 54.5 & 60.0 & 78.9 & 90.9 & 100.0 & 86.4 \\
    GLM-4.5V-106B & \textbf{82.9} & \textbf{77.6} & \textbf{77.0} & 83.5 & \textbf{80.1} & & 78.6 & 81.2 & 75.0 & \textbf{70.0} & 69.2 & 76.9 & 84.2 & \textbf{100.0} & 92.9 & 81.8 \\
    GUI-OWL-7B & 68.4 & 72.4 & 76.8 & 63.2 & 69.9 & & 66.7 & 66.7 & 64.7 & 55.6 & 52.6 & 71.4 & 76.2 & \textbf{100.0} & 84.2 & 94.1 \\
    GUI-OWL-32B & 75.0 & 71.4 & 71.2 & 75.2 & 73.2 & & 70.0 & 65.0 & 68.8 & 60.0 & \textbf{90.0} & \textbf{81.2} & 77.8 & 83.3 & 85.0 & \textbf{100.0} \\
    SE-WSM-7B & 70.0 & 52.2 & 20.1 & \textbf{91.0} & 54.8 & & \textbf{100.0} & 57.1 & 62.5 & 44.4 & 66.7 & 56.5 & \textbf{88.9} & 61.9 & \textbf{100.0} & 55.9 \\
    \midrule
    \multicolumn{17}{c}{\textit{voting-majority}} \\
    \midrule
    G106-s 2runs & 84.3 & 70.7 & 65.5 & 87.2 & 76.1 & & 76.9 & 76.5 & 71.4 & 58.3 & \textbf{70.0} & 68.8 & 88.2 & 92.3 & \textbf{100.0} & 76.0 \\
    Q32-s + G106-s & \textbf{90.1} & 68.5 & 59.0 & \textbf{93.2} & 75.7 & & \textbf{90.9} & 78.9 & \textbf{87.5} & 55.6 & 62.5 & 61.1 & \textbf{92.3} & 76.5 & 92.3 & 78.3 \\
    \makecell[l]{Q32-s + G106-s \\+ G106-z} & 81.6 & \textbf{84.8} & \textbf{86.3} & 79.7 & \textbf{83.1} & & 80.0 & \textbf{86.7} & 73.7 & \textbf{85.7} & 68.8 & \textbf{90.0} & 72.7 & \textbf{100.0} & 93.8 & \textbf{90.0} \\
    \midrule
    \multicolumn{17}{c}{\textit{voting-strict\_unanimous}} \\
    \midrule
    G106-s 2runs & 84.3 & 82.3 & \textbf{65.5} & \textbf{76.7} & \textbf{71.0} & & 76.9 & 86.7 & 71.4 & 75.0 & \textbf{70.0} & 76.9 & 88.2 & \textbf{100.0} & \textbf{100.0} & 90.0 \\
    Q32-s + G106-s & \textbf{90.1} & 84.2 & 59.0 & 72.2 & 65.4 & & \textbf{90.9} & 85.7 & \textbf{87.5} & 83.3 & 62.5 & \textbf{90.0} & \textbf{92.3} & \textbf{100.0} & 92.3 & 89.5 \\
    \makecell[l]{Q32-s + G106-s \\+ G106-z (\textbf{UPE})} & 89.8 & \textbf{93.3} & 56.8 & 63.2 & 59.9 & & \textbf{90.9} & \textbf{100.0} & \textbf{87.5} & \textbf{100.0} & 57.1 & \textbf{90.0} & \textbf{92.3} & \textbf{100.0} & 92.3 & \textbf{100.0} \\
    \bottomrule
    \end{tabular}
    \caption{Performance comparison of outcome reward models (ORM) 
    across different vision-language models, prompt configurations, 
    and task categories. 
    Results show precision (P), negative predictive value (NPV), recall (R), and specificity (S) for trajectory success evaluation 
    under two prompt configurations (\textit{zerogui} and \textit{sewsm}) and two voting strategies (\textit{voting-majority} and \textit{voting-strict\_unanimous}). 
    In model names, ``-z'' denotes \textit{zerogui} prompt and ``-s'' denotes \textit{sewsm} prompt; 
    ``Q32'' denotes Qwen2.5VL-32B and ``G106'' denotes GLM-4.5V-106B.
    Due to space constraints, complete results across all task categories are provided in Table \ref{tab:appendix_reward_model_performance}.} 
    \label{tab:reward_model_performance}
\end{table*}

\begin{table*}[t]
    \centering
    \scriptsize  % 改为更小的字体
    \setlength{\tabcolsep}{3.8pt}  % 减小列间距
    \begin{tabular}{lccccc|cc|cc|cc|cc|cc}
    \toprule
    \multirow{2}{*}{Reward Model} & \multicolumn{5}{c|}{Overall} & \multicolumn{2}{c|}{vscode} & \multicolumn{2}{c|}{gimp} & \multicolumn{2}{c|}{writer} & \multicolumn{2}{c|}{chrome} & \multicolumn{2}{c}{multi\_apps} \\
    & P & NPV & R & S & OA & P & NPV & P & NPV & P & NPV & P & NPV & P & NPV \\
    \midrule
    \multicolumn{16}{c}{\textit{opencua\_reflector}} \\
    \midrule
    Qwen2.5VL-7B & 54.4 & 49.4 & 53.8 & 50.0 & 52.0 & 53.6 & 42.9 & 59.1 & 50.0 & 54.5 & 50.0 & 44.0 & 43.5 & 63.2 & 57.7 \\
    Qwen2.5VL-32B & 60.3 & 64.8 & 79.8 & 41.5 & 61.7 & 58.3 & 66.7 & \textbf{76.2} & 73.3 & 60.0 & \textbf{66.7} & 57.1 & \textbf{69.2} & 62.5 & 76.9 \\
    Qwen2.5VL-72B & 58.5 & 64.8 & 83.0 & 34.8 & 60.1 & 64.5 & 72.7 & 73.1 & 90.0 & 54.2 & 60.0 & 52.4 & 66.7 & 56.2 & 61.5 \\
    GLM-4.5V-106B & 64.0 & \textbf{78.5} & \textbf{89.0} & 44.5 & \textbf{67.9} & \textbf{66.7} & \textbf{88.9} & 74.1 & \textbf{100.0} & 57.9 & 60.0 & 57.1 & \textbf{69.2} & \textbf{71.0} & \textbf{92.9} \\
    GUI-OWL-7B & \textbf{64.6} & 61.8 & 67.0 & \textbf{59.1} & 63.3 & 63.6 & 55.0 & 66.7 & 60.0 & \textbf{62.5} & 61.5 & 55.2 & 57.9 & 66.7 & 62.5 \\
    GUI-OWL-32B & 61.6 & 61.5 & 71.4 & 50.6 & 61.6 & 62.5 & 70.0 & 66.7 & 66.7 & 56.2 & 53.8 & \textbf{59.4} & 68.8 & 70.0 & 64.0 \\
    \midrule
    \multicolumn{16}{c}{\textit{sewsm}} \\
    \midrule
    Qwen2.5VL-7B & 56.7 & 54.2 & 67.0 & 43.3 & 55.8 & 62.5 & 55.6 & 50.0 & 40.9 & 56.2 & 53.8 & 57.1 & 69.2 & 55.6 & 55.6 \\
    Qwen2.5VL-32B & 65.3 & 62.4 & 67.8 & 59.8 & 64.0 & 74.1 & 80.0 & 76.5 & 63.2 & 69.2 & 62.5 & 71.4 & 66.7 & \textbf{66.7} & 66.7 \\
    Qwen2.5VL-72B & 58.7 & 67.1 & 85.2 & 33.5 & 60.7 & 60.0 & 71.4 & 72.7 & 71.4 & 58.3 & 80.0 & 56.4 & 77.8 & 56.2 & 61.5 \\
    GLM-4.5V-106B & \textbf{69.5} & 64.2 & 66.1 & \textbf{67.7} & \textbf{66.9} & 68.8 & \textbf{90.0} & \textbf{81.2} & 65.0 & \textbf{72.7} & 61.1 & \textbf{80.0} & 71.4 & 63.6 & 60.9 \\
    GUI-OWL-7B & 56.6 & 64.2 & 86.8 & 26.2 & 58.1 & 59.4 & 60.0 & 59.3 & 55.6 & 54.2 & 60.0 & 53.5 & \textbf{80.0} & 52.6 & 57.1 \\
    GUI-OWL-32B & 57.4 & \textbf{68.8} & \textbf{89.0} & 26.8 & 59.5 & 57.9 & 75.0 & 64.3 & \textbf{75.0} & 70.0 & \textbf{88.9} & 53.8 & 66.7 & 55.0 & \textbf{80.0} \\
    SE-WSM-7B & 58.7 & 52.0 & 48.4 & 62.2 & 54.9 & \textbf{76.2} & 66.7 & 58.3 & 45.8 & 70.0 & 57.9 & 60.0 & 57.1 & 57.9 & 53.8 \\
    \midrule
    \multicolumn{16}{c}{\textit{voting-majority}} \\
    \midrule
    G106-s 2runs & 73.6 & 62.8 & 59.6 & 76.2 & 67.4 & 72.4 & 84.6 & 81.2 & 65.0 & 72.7 & 61.1 & \textbf{85.7} & 64.7 & 68.8 & 58.6 \\
    Q32-s + G106-s & \textbf{75.6} & 60.5 & 52.5 & \textbf{81.1} & 66.0 & \textbf{79.2} & 77.8 & \textbf{100.0} & 64.0 & \textbf{85.7} & 59.1 & 83.3 & 61.1 & \textbf{76.5} & 64.3 \\
    \makecell[l]{Q32-s + G106-s \\+ G106-o} & 68.4 & \textbf{71.5} & \textbf{78.6} & 59.8 & \textbf{69.7} & 71.9 & \textbf{100.0} & 83.3 & \textbf{72.2} & 68.8 & \textbf{69.2} & 80.0 & \textbf{82.6} & 70.8 & \textbf{71.4} \\
    \midrule
    \multicolumn{16}{c}{\textit{voting-strict\_unanimous}} \\
    \midrule
    G106-s 2runs & 73.6 & 64.2 & \textbf{59.6} & \textbf{57.9} & \textbf{58.8} & 72.4 & 88.9 & 81.2 & 68.4 & 72.7 & 56.2 & 85.7 & 75.0 & 68.8 & 59.1 \\
    Q32-s + G106-s & 75.6 & 69.1 & 52.5 & 46.3 & 49.6 & 79.2 & \textbf{100.0} & \textbf{100.0} & 64.3 & \textbf{85.7} & \textbf{66.7} & 83.3 & 84.2 & 76.5 & 62.5 \\
    \makecell[l]{Q32-s + G106-s \\+ G106-o (\textbf{UPE})} & \textbf{81.7} & \textbf{85.1} & 48.9 & 24.4 & 37.3 & \textbf{81.8} & \textbf{100.0} & \textbf{100.0} & \textbf{100.0} & 83.3 & 62.5 & \textbf{87.5} & \textbf{100.0} & \textbf{86.7} & \textbf{85.7} \\
    \bottomrule
    \end{tabular}
    \caption{Evaluation results of process reward models (PRM) 
    for step-level correctness assessment across various vision-language models 
    and prompt configurations. 
    The table presents precision (P), negative predictive value (NPV), recall (R), and specificity (S) metrics 
    under two prompt configurations (\textit{opencua\_reflector} and \textit{sewsm}) and two voting strategies (\textit{voting-majority} and \textit{voting-strict\_unanimous}). 
    In model names, ``Q32'' denotes Qwen2.5VL-32B and ``G106'' denotes GLM-4.5V-106B.
    Due to space constraints, complete results across all task categories are provided in Table \ref{tab:appendix_action_reward_model_performance}.}
    \label{tab:action_reward_model_performance}
\end{table*} 

\subsection{Effect of VLM Selection}
% 1. 参数量和训练质量对reward model的性能有着全面的影响。
% 1）无论是通用VLM还是CUA VLM、以及专用数据训练的reward model，
% 所有的7B模型表现在不同指标上都都落后于32B以上的模型。
% 尽管专门的CUA训练使得7B模型的reward prediction能力
% 有所提升（GUI-OWL-7B vs Qwen2.5VL-7B），
% 但仍旧落后于32B以上的模型。
% 2）Surprisingly, Qwen2.5VL-72B underperforms compared to the 32B variant, 
% with particularly pronounced differences in action-level evaluation. 
% A possible explanation is that Qwen2.5VL-32B underwent 
% additional reinforcement learning training 
% compared to the 72B model\footnote{https://huggingface.co/Qwen/Qwen2.5-VL-32B-Instruct}, resulting in enhanced generalization capabilities for reward modeling tasks.
\textbf{Model size and training quality comprehensively impact reward model performance.}
1) Across all evaluated models—including general VLMs, 
specialized CUA models VLMs, and specialized CUA reward models—7B variants 
consistently underperform their 32B+ counterparts across different metrics.
While CUA-specific training 
enhances the reward prediction capabilities of 7B models 
(e.g., GUI-OWL-7B vs. Qwen2.5VL-7B in Table \ref{tab:reward_model_performance} and Table \ref{tab:action_reward_model_performance}), 
they still lag behind larger models.
2) Surprisingly, Qwen2.5VL-72B underperforms compared to the 32B variant, 
with particularly notable differences in action-level evaluation.
A possible explanation is that Qwen2.5VL-32B received additional RL training 
compared to the 72B model\footnote{https://huggingface.co/Qwen/Qwen2.5-VL-32B-Instruct}, 
which improved its reasoning capabilities and consequently led to stronger generalization for CUA reward prediction.

% 2. 
% 表象：Qwen2.5VL-32B和GLM-4.5V-106B视觉推理模型表现最佳
% 分析：
% Qwen2.5VL-32B vs GUI-OWL-32B，后者是从前者后训练而来
% 但是在任何prompt setting下，后者都比前者表现更差。
% 通过对response的具体比较，我们发现，GUI-OWL-32B的reasoning过程比Qwen2.5VL-32B的短很多，
% 这意味着在post-training中虽然混合了两部分的general reasoning数据，
% 但还是丧失了一定的reasoning能力。
% GLM-4.5V-106B胜过其它所有模型也印证了这一点。
% 关键性结论：视觉推理能力是CUA reward model的核心要素，专门化训练必须在保持推理能力的前提下进行。"
\textbf{Visual reasoning capability is the core element of CUA reward models. }
GUI-OWL-32B, despite being post-trained from Qwen2.5VL-32B 
specifically for CUA tasks, 
consistently underperforms its base model across all prompt configurations.
Through detailed response analysis, 
we observe that GUI-OWL-32B produces significantly shorter reasoning processes 
compared to Qwen2.5VL-32B, 
indicating that despite incorporating general reasoning data during post-training, 
the model still experiences some degradation in reasoning capabilities.
The superior performance of GLM-4.5V-106B over all other models 
further corroborates this finding, 
as it maintains the strongest visual reasoning abilities 
among the evaluated models. 

% 3. 
% 表象：CUA模型的performance悖论，GUI-OWL是从Qwen2.5VL微调而来
% GUI-OWL-7B比Qwen2.5VL-7B提升非常多，
% 但是GUI-OWL-32B比Qwen2.5VL-32B有明显退步。
% 分析：
% Qwen2.5VL-7B的推理能力本身就弱，因而CUA训练带来的增益超过了推理能力削弱的负面影响
% 但是Qwen2.5VL-32B推理能力较强，推理能力削弱的负面影响掩盖了CUA训练带来的增益。
% 关键性结论：
% 针对CUA的policy training有益于CUA reward评估能力的提升，但前提是不能损害Reasoning能力

% 放到末尾的开放性思考：这对现有的CUA训练范式和CUA RM的训练范式都有启发，
% 二者的训练应该融合在一起或许是更有用的范式
\textbf{CUA policy training benefits reward evaluation capabilities, 
but only when reasoning abilities are preserved.}
GUI-OWL exhibits a performance paradox: 
GUI-OWL-7B significantly outperforms Qwen2.5VL-7B, 
while GUI-OWL-32B shows notable degradation compared to Qwen2.5VL-32B.
This counterintuitive pattern stems from 
the differential impact of CUA specialization 
on models with varying baseline reasoning capabilities.
For Qwen2.5VL-7B, which possesses inherently weak reasoning abilities, 
the benefits gained from CUA training 
outweigh the negative effects of reasoning capability degradation.
Conversely, for Qwen2.5VL-32B with strong baseline reasoning capabilities, 
the negative impact of reasoning degradation 
overshadows the gains from CUA training.
This finding suggests that CUA policy training 
can enhance reward evaluation performance, 
but only when the specialization 
process preserves the model's fundamental reasoning abilities.
These implications inform both CUA policy training and reward model development, 
suggesting that incorporating sufficient high-quality general reasoning data 
during CUA training is essential 
to maintain the model's reasoning capabilities 
and thereby ensure effective reward evaluation performance.

% 4. 
% 表象：SE-WSM的性能崩溃。
% 分析：数据覆盖面太窄
% 其训练数据限定在43个feasible chrome tasks in OSWorld 的860条轨迹，
% 这或许足以使得其在
% 都是web任务的AgentRewardBench\cite{lu2025agentrewardbench}上表现出色，
% 但是却难以泛化到task category全面的CUARewardBench。，
% 关键性结论：specialized CUA RM应当注重训练数据的多样性
% 末尾补充一下自夸：
% 正是CUARewardBench的测试轨迹全面的task-category覆盖
% 和policy model多样性所带来的挑战性，使其足够成为CUA RM的试金石。
\textbf{Specialized CUA reward models 
require diverse training data to achieve effective generalization.}
SE-WSM \cite{seagent}, a specialized CUA reward model fine-tuned from Qwen2.5VL-7B, shows overall performance on CUARewardBench that is comparable to, but slightly worse than, its Qwen2.5VL-7B base model.
This underperformance likely stems from its narrowly scoped training data coverage,
which comprises only 860 trajectories from 43 Chrome tasks.
While this narrow scope may suffice for web-only benchmarks 
like AgentRewardBench~\cite{lu2025agentrewardbench},
it proves inadequate for generalizing 
to the comprehensive task categories covered in our CUARewardBench.
The comprehensive task-category coverage 
and policy model diversity in CUARewardBench
successfully expose limitations that remain hidden in more limited evaluation settings,
demonstrating its effectiveness as a rigorous testbed for CUA RM capabilities.

%1. \textbf{相比于模型对评估的影响是P和R的整体提升，prompt则是影响着P和R的trade-off via 标准的宽松程度}
% As shown in Tables~\ref{tab:reward_model_performance} 
% 以Qwen2.5-32B为例，ORM的setting下，采用\textit{sewsm}和\textit{zerogui}的prompt，F1和OA接近
% 采用\textit{sewsm}的prompt，precision比zerogui prompt 下高9.1
% recall却低11.5。
% 同样的情况也发生在PRM的setting下，
% as shown in Tables~\ref{tab:action_reward_model_performance},
% 采用\textit{sewsm}的prompt，precision比\textit{opencua_reflector} prompt 高9.2个百分点，
% 但是recall却低13.7个百分点。

%2. \textbf{Prompt \textit{sewsm} vs prompt \textit{zerogui}.} 
% 通过对prompt和response的对比分析，我们发现，造成二者差异的原因在于\textit{sewsm}有着
% 比\textit{zerogui}更加严格的成功标准。
% \textit{sewsm}要求VLM从更多的维度来验证当前轨迹的合理性，
% 包括轨迹正确性、冗余步骤、first error step、correct action suggestion
% 引入的这些维度自然就造成了成功的轨迹有更大的可能因为某些方面的不完美而被误杀。
% 与之相反，\textit{zerogui}的成功标准则更加宽松，只要求VLM单调地判断轨迹成功与失败，
% 这就造成了某些失败轨迹容易通过表面的合理来骗过VLM。

%3.\textbf{ Prompt \textit{sewsm} vs prompt \textit{opencua_reflector}. }
% Prompt \textit{opencua_reflector} 相比prompt \textit{sewsm}最大的区别在于，
% 仅仅接受到当前步骤为止的truncated trajectory。
% 这就造成了前者缺乏后向视角，仅能观察到当前action对当下状态的影响，
% 无法获得全局视野来判断对后续步骤乃至整个任务的影响。
% 这使得前者在评估轨迹的正确性上，容易轻信某些具备迷惑性的bad action，将其判定为good action。
% 与之相反，\textit{sewsm}的prompt则能够接收到完整的trajectory，
% 这使得RM自然拥有了更严格的约束来action的正确性。
\subsection{Impact of Prompt Templates}
\label{sec:prompt_templates_impact}
\textbf{Prompt templates primarily influence P-NPV trade-offs rather than overall performance improvements.}
As shown in Tables~\ref{tab:reward_model_performance} and \ref{tab:action_reward_model_performance}, 
different prompt templates create distinct evaluation standards that affect the precision-recall balance.
Taking Qwen2.5VL-32B as an example, in ORM settings, 
the \textit{sewsm} prompt achieves 9.1 percentage points higher precision than \textit{zerogui}, 
but suffers an 4.4 percentage point drop in NPV.
This trade-off pattern is consistently observed in PRM settings, 
where \textit{sewsm} outperforms \textit{opencua\_reflector} by 5.0 percentage points in precision 
while losing 2.4 percentage points in NPV.

\textbf{SE-WSM prompt enforces stricter trajectory success criteria compared to ZeroGUI.}
Through comparative analysis of prompts and model responses, 
we find that \textit{sewsm} employs stricter success standards than \textit{zerogui}.
The \textit{sewsm} template requires VLMs to verify trajectory reasonableness across multiple dimensions,
including trajectory correctness, redundant steps, first error identification, and correct action suggestions.
These additional evaluation dimensions naturally increase the likelihood 
that successful trajectories may be incorrectly rejected due to minor imperfections.
Conversely, \textit{zerogui} adopts relaxed criteria, 
requiring only binary trajectory success determination,
which allows some failed trajectories to deceive VLMs through superficial reasonableness.

\textbf{OpenCUA reflector adopts more relaxed action evaluation criteria compared to SE-WSM.}
The key distinction between \textit{opencua\_reflector} and \textit{sewsm} prompts 
lies in their temporal scope: 
\textit{opencua\_reflector} receives only truncated trajectories up to the current step,
limiting its perspective to immediate action effects on current states
without global visibility into subsequent steps or overall task impact.
This constraint makes the former susceptible to deceptive bad actions 
that appear reasonable in isolation but prove detrimental to task completion.
In contrast, \textit{sewsm} processes complete trajectories,
naturally providing stricter constraints for action correctness evaluation 
through comprehensive temporal context.

\subsection{Verification Difficulty in CUA}
\textbf{Reward model remains unreliable for both trajectory- and step-level CUA assessments.}
Comparing CUA success rates in Table~\ref{tab:agent_performance} with ORM overall accuracy in Table~\ref{tab:reward_model_performance}, we observe that while trajectory-level verification asymmetry \cite{wei2024asymmetry} remains evident, even the best-performing ORM achieves only 82.9\% precision and 80.1\% overall accuracy, indicating substantial room for improvement.
Similarly, Table~\ref{tab:action_reward_model_performance} shows that step-level verification also presents considerable challenges, with reward models demonstrating limited effectiveness in providing reliable step-by-step guidance.
These findings reveal that both trajectory-level ORMs 
and step-level PRMs fall short of ideal performance standards, 
collectively limiting their capacity 
to provide reliable supervision signals for CUA training.

\subsection{Ensemble Methods}
\label{sec:ensemble_methods}

% 1. Strict Unanimous Voting: 牺牲sample efficiency提高reward reliability。
% ZeroGUI\cite{zerogui}采用的是Majority Voting，但是实验结果表明，
% 这种Voting虽然能提升precion，
% 但是却会导致NPV的大幅降低（Q32-s + G106-s vs single G106-s 
% in Table \ref{tab:reward_model_performance} and Table \ref{tab:action_reward_model_performance}）。
% 考虑到在实际应用中，reward reliability is more critical than sample efficiency，
% 我们针对性地提出了一种新的投票策略，strict unanimous voting。
% 一方面，它和传统的unanimous voting \cite{kuncheva2014combining} 一样，要求投票一致性；
% 另一方面，它要求对正、负预测结果都要求一致性，假如不能达成一致，就放弃该样本。
% 它和Majority Voting的区别如表\ref{tab:voting_strategies}所示。
% As show in Table \ref{tab:reward_model_performance} and Table \ref{tab:action_reward_model_performance}，
% 采用strict unanimous voting策略，无论是ORM还是PRM，
% P和NPV都有了大幅的提升，Recall和Specificity的跌幅虽然也很大，但仍在可接受的范围内。

% 2. Prompt-Template Ensemble：利用不同prompt-template对P和NPV的trade-off差异增强投票组合。
% 正如\ref{sec:prompt_templates_impact}提到的，在不同prompt-template下，
% 对于P和NPV的trade-off差异是不同的。
% 我们可以利用这一点，在投票组合中引入不同的prompt-template组合，
% 实验结果显示，对于ORM，这能带来NPV的进一步提升；
% 对于PRM，则是P和NPV的均被大幅提升。
Building upon the insights from previous analyses, 
we propose \textbf{Unanimous Prompt Ensemble (UPE)}, 
a novel ensemble approach that 
significantly enhances reward model reliability for CUA tasks. 
UPE integrates two complementary strategies: 
(1) a strict unanimous voting mechanism 
that prioritizes prediction reliability over sample efficiency, 
and (2) strategic prompt-template configurations 
that leverage the complementary P-NPV trade-offs 
identified in Section~\ref{sec:prompt_templates_impact}. 
Together, these components enable substantial improvements 
in both precision and negative predictive value, 
which are critical metrics for ensuring reliable reward signals 
in reinforcement learning applications.

\textbf{Strict Unanimous Voting.}
While ZeroGUI~\cite{zerogui} employs majority voting, 
our experiments reveal a critical limitation: 
it improves precision but substantially reduces NPV 
(e.g., Q32-s + G106-s vs. G106-s in Tables~\ref{tab:reward_model_performance} 
and \ref{tab:action_reward_model_performance}). 
In CUA training, reward reliability is more critical than sample efficiency. 
Reduced sample efficiency can be compensated by increased sampling, 
but unreliable rewards directly compromise RL training quality. 
Therefore, we introduce \textbf{strict unanimous voting}.
Extending traditional unanimous voting~\cite{kuncheva2014combining}, 
our strategy requires consensus on both positive and negative predictions: 
a sample is classified only when all ensemble members unanimously agree; 
otherwise, it is abstained. 
Table~\ref{tab:voting_strategies} illustrates the distinction 
with concrete voting scenarios, 
comparing how strict unanimous voting 
and majority voting produce different decisions under identical ensemble configurations. 
As shown in Table~\ref{tab:reward_model_performance} and \ref{tab:action_reward_model_performance},
this approach substantially improves both precision and NPV 
for ORM and PRM. 
While recall and specificity decrease, 
they remain acceptable—a favorable trade-off 
when reliability outweighs coverage.
\begin{table}[h]
    \centering
    \footnotesize
    \begin{tabular}{lcc}
    \toprule
    \textbf{Voting Scenario} & \textbf{Majority Voting} & \textbf{Strict Unanimous} \\
    \midrule
    2 Pos., 0 Neg. & Positive $\checkmark$ & Positive $\checkmark$ \\
    \midrule
    1 Pos., 1 Neg. & Negative $\times$ & No Prediction $\triangle$ \\
    \midrule
    0 Pos., 2 Neg. & Negative $\times$ & Negative $\times$ \\
    \bottomrule
    \end{tabular}
    \caption{Comparison between majority voting and strict unanimous voting strategies. $\checkmark$ indicates positive prediction, $\times$ indicates negative prediction, and $\triangle$ indicates no prediction when consensus cannot be reached.}
    \label{tab:voting_strategies}
\end{table} 

\textbf{Prompt-Template Ensemble.}
As revealed in Section~\ref{sec:prompt_templates_impact}, 
different prompt templates 
exhibit distinct P-NPV trade-off characteristics. 
We exploit this complementarity 
by strategically combining models configured 
with diverse prompt templates within our voting ensemble. 
As shown in Table~\ref{tab:reward_model_performance},
Q32-s + G106-s +G106-z outperforms Q32-s + G106-s: while precision decreases slightly by 0.3 percentage points, NPV increases by 9.1 percentage points. 
This demonstrate that for ORM, 
this heterogeneous prompt configuration 
yields further NPV improvements beyond those 
achieved by strict unanimous voting strategy alone. 
For PRM, the benefits are even more pronounced (Q32-s + G106-s +G106-o vs. Q32-s + G106-s in Table~\ref{tab:action_reward_model_performance}), 
with substantial simultaneous gains in both precision and NPV. 
This synergy between strict unanimous voting and prompt diversity 
establishes UPE as an effective method 
for enhancing reward model reliability in CUA evaluation.

\section{Error Analysis}
\label{sec:error_analysis}
%我们对Table\ref{sec:reward_performance}中表现最好的GLM-4.5V-106B
% 的ORM的53个badcase进行分析，并总结其错误模式。
% As shown in Table~\ref{tab:error_analysis}, 
% 按照占比大小排序，错误模式依次为推理错误、视觉理解错误、知识缺陷、RM局限性。
% 下面我们详细分析每种错误模式。（每个的段落标题用\textfb）
This section provides a comprehensive microscopic analysis of error patterns, 
examining specific failure modes,
and the fundamental limitations that constrain current reward model effectiveness.
We analyze 53 failure cases from GLM-4.5V-106B, 
the best-performing model in ORM evaluation, 
to identify systematic error patterns. 
As shown in Table~\ref{tab:error_analysis}, 
we categorize errors by frequency: reasoning errors (35.8\%), visual understanding errors (30.2\%), action understanding errors (17.0\%), knowledge deficiency (15.1\%), and inherent RM limitations (1.9\%). We examine each category in detail below.

\begin{table}[h]
    \centering
    \footnotesize  % 改为更小的字体
    \begin{tabular}{lcc}
    \toprule
    Error Category & Count & Percentage (\%) \\
    \midrule
    Reasoning Error & 19 & 35.8 \\
    Visual Understanding Error & 16 & 30.2 \\
    Action Understanding Error & 9 & 17.0 \\
    Knowledge Deficiency & 8 & 15.1 \\
    Inherent RM Limitation & 1 & 1.9 \\
    \midrule
    \textbf{Total} & \textbf{53} & \textbf{100} \\
    \bottomrule
    \end{tabular}
    \caption{Distribution of error modes of GLM-4.5V-106B as ORM using \textit{sewsm} prompt.}
    \label{tab:error_analysis}
\end{table}

% 1 视觉理解错误(30\%)
% RM经常错误理解screenshot中的视觉信息，导致对computer state得出错误的估计。
% 例如，agent正在执行任务"... adding strike-through sign on the line ..."，
% 但是选中并添加strike-through漏了末尾的几个字母。
% RM没能察觉出agent的这一错误，因此判定任务成功。
\textbf{Visual Understanding Errors (30.2\%).}
Reward models frequently misinterpret visual information in screenshots, leading to incorrect assessments of computer states. For instance, when an agent executes the task "adding strike-through sign on the line," it successfully selects and applies strike-through formatting but misses the final few characters. The reward model fails to detect this incomplete execution and incorrectly judges the trajectory as successful.

% 2 动作理解错误(17%)
% 在SE-WSM的prompt下，RM根据历史全部截图来推断agent的动作，进而判断轨迹的成功与否。
% 然而，RM很容易从相邻截图中推导出错误的动作。
% 例如，agent在对话框中点击了OK按钮，但是RM却误以为agent没有完成确认操作，
% 因此没有完成任务。
% 一个简单的解决方式是像OpenCUA reflector那样在截图中标注坐标marker，
% 因此，相比之下，采用opencua_reflector的prompt的RM极少犯下这样的错误。
\textbf{Action Understanding Errors (17.0\%).}
Under the SE-WSM prompt configuration, reward models infer agent actions from consecutive screenshots to evaluate trajectory success. However, models often derive incorrect actions from adjacent frames. For example, when an agent clicks an "OK" button in a dialog box, the reward model mistakenly believes the agent failed to complete the confirmation operation, leading to an incorrect failure assessment. A straightforward solution involves incorporating coordinate markers in screenshots, as implemented in the OpenCUA reflector approach, which significantly reduces such errors.

% 3 知识缺陷(15%)
% Agent很容易因为缺乏软件操作的相关知识而任务失败，
% 同样的，RM也经常因为缺乏相关知识而没法建立起正确的任务成功标准。
% 例如，agent的任务是"...enlarge the text on my screen ..."
% 但是却搞成了放大整个屏幕而非text font
% RM不知道ubuntu系统设置提供这样不同的选项，因此判定任务成功
\textbf{Knowledge Deficiency (15.1\%).}
Just as agents frequently fail due to insufficient software operation knowledge, reward models often lack domain-specific knowledge necessary to establish correct task success criteria. For instance, when an agent's task is to "enlarge the text on my screen," the agent incorrectly magnifies the entire screen rather than adjusting text font size. The reward model, unaware that Ubuntu system settings provide distinct options for these operations, incorrectly judges the trajectory as successful.

% 4 推理错误(36%)
% 很多时候，即便RM正确理解了screenshot的视觉元素以及agent执行的action，
% 并且不缺乏相关知识，RM仍然会在组织这些信息的过程中犯下逻辑错误。
% 例如，agent的任务是"... set the decimal separator as a comma (,) ..."
% 并且正确执行成功。
% RM先是认可了agent的完成了设置，但是又在一番冗长的思考中重新收集信息并推翻了原有的结论。
\textbf{Reasoning Errors (35.8\%).}
Even when reward models correctly understand visual elements and agent actions while possessing relevant knowledge, they frequently commit logical errors during information synthesis. For example, when an agent successfully completes the task "set the decimal separator as a comma (,)," the reward model initially acknowledges the correct configuration but subsequently engages in convoluted reasoning that leads to overturning its original correct conclusion.

% 5 RM固有缺陷(1.9%)
% 还有一类偶发性的错误占比极小，但却体现了当前VLM-based RM的固有缺陷，
% 截图只是对computer state的部分观察。
% 例如agent的任务是“... use GIMP to compress the image to under 600KB ...”
% 并且成功完了任务。
% 但是截图中并没有反馈出压缩后的图片大小，因此RM没有证据来判定任务是否被完成。
% 这类错误启发我们，RM和scripted-based verifier或许可以作为相互补充以实现更鲁邦的reward estimation。
\textbf{Inherent RM Limitations (1.9\%).}
A small but significant category of errors reveals fundamental limitations of VLM-based reward models: screenshots provide only partial observations of computer states. For instance, when an agent successfully completes the task "use GIMP to compress the image to under 600KB," the screenshot lacks visual feedback about the compressed file size, leaving the reward model without evidence to verify task completion. This limitation suggests that reward models and script-based verifiers could serve as complementary approaches for more robust reward estimation.

%RM vs scrip-based traj evaluator
%1) RM的局限性分析：只能和执行agent共享视野，但是有的时候执行agent的轨迹是错误的，screenshot没有验证所需的信息。
%2) scrip-based traj evaluator相比于RM的局限性

\section{Limitations and Future Directions}

While CUARewardBench provides a rigorous benchmark for CUA reward model evaluation, our work has several important limitations that warrant discussion:

\textbf{Limited Practical Validation of UPE}
Although we propose the UPE method 
and demonstrate its effectiveness 
in improving reward reliability (precision and NPV), 
we have not validated it in actual reinforcement learning training loops. 
This leaves several critical questions unanswered: 
(1) Could the trade-off between sample efficiency 
and reward reliability observed in our benchmark be maintained 
in the rollouts of actual RL training? 
(2) Could the samples filtered by UPE exhibit systematic biases, 
potentially discarding high-value training examples? 
(3) How does UPE perform 
across different training stages and policy distributions? 
These questions require extensive RL training experiments to address.

\textbf{Benchmark Scale and Scenario Coverage}
While CUARewardBench establishes rigorous annotation standards, 
two key limitations constrain its scope:
(1) \textit{Limited scale}: With 272 trajectory annotations and 346 step-level annotations, 
the current benchmark may not fully capture long-tail failure modes or diverse agent strategies.
(2) \textit{Task distribution gap}: All tasks are sampled from OSWorld \cite{osworld}, 
which focuses primarily on limited application scenarios. 
This creates a gap with real-world computer use, 
where more diverse and complex workflows are prevalent.

\section{Conclusions}
This paper presents a systematic investigation into computer-using agent reward models through benchmark construction, empirical analysis, and method development. We introduce CUARewardBench, the first comprehensive benchmark for evaluating reward models on computer-using agents, comprising 272 trajectory annotations and 346 step-level annotations across 10 software categories. Through systematic evaluation of 7 vision-language models with 3 prompt templates, we reveal critical insights into current CUA RM capabilities and limitations. Our key findings include that: (1) model size and training quality comprehensively impact reward model performance; (2) visual reasoning capability is the core element of CUA reward models, with general VLMs outperforming specialized CUA models; (3) prompt templates primarily influence precision-recall trade-offs rather than overall performance improvements; and (4) both trajectory-level and step-level verification face significant challenges, with action-level evaluation proving more difficult than trajectory-level assessment. Error analysis reveals that reasoning errors (35.8\%) and visual understanding errors (30.2\%) constitute the primary failure modes. Building upon these insights, we propose Unanimous Prompt Ensemble (UPE), a novel ensemble method that significantly enhances reward model reliability through strict unanimous voting and strategic prompt-template configurations. UPE achieves 89.8\% precision and 93.3\% NPV for ORM, and 81.7\% precision and 85.1\% NPV for PRM, substantially outperforming single VLMs and traditional ensemble approaches.

Our work establishes both a rigorous evaluation framework and an immediately deployable solution for the community. CUARewardBench provides a standardized testbed for advancing computer-using agent evaluation, while UPE offers a practical method for enhancing reward model reliability in CUA training pipelines. Together, these contributions lay the foundation for developing more reliable reward models to support large-scale CUA training and deployment.

\clearpage
\newpage
\section*{Contributions}
\paragraph{Authors}
Haojia Lin\textsuperscript{\rm 1*}\quad 
Xiaoyu Tan\textsuperscript{\rm 1*}\quad 
Yulei Qin\textsuperscript{\rm 1*}\quad 
Zihan Xu\textsuperscript{\rm 1}\quad 
Yuchen Shi\textsuperscript{\rm 1}\quad 
Zongyi Li\textsuperscript{\rm 1}\quad 
Gang Li\textsuperscript{\rm 1}\quad  
Shaofei Cai\textsuperscript{\rm 1,2}\quad 
Siqi Cai\textsuperscript{\rm 1}\quad 
Chaoyou Fu\textsuperscript{\rm 3}\quad 
Ke Li\textsuperscript{\rm 1}\quad 
Xing Sun\textsuperscript{\rm 1}

\paragraph{Affiliations}
\textsuperscript{\rm 1}Tencent Youtu Lab\quad 
\textsuperscript{\rm 2}Peking University\quad 
\textsuperscript{\rm 3}Nanjing University

\paragraph{$^*$Equal Contributions}
Haojia Lin\quad  
Xiaoyu Tan\quad 
Yulei Qin

\paragraph{Acknowledgments}
We greatly thank the OSWorld \cite{osworld,osworld_verified} community for open-sourcing the CUA tasks and diverse CUA trajectories.

\clearpage
\newpage
\setcitestyle{numbers,square}
%% \setcitestyle{square,numbers,comma}
%% \bibliographystyle{unsrt}
%% \bibliographystyle{plainnat}

\bibliography{youtu_bib}

\clearpage
\newpage
\section{Appendix}

\subsection{Related Work}
\label{sec:related_work}

\textbf{Computer-Use Agents. }
Computer-use agent approaches can be broadly categorized into three methodological paradigms. 
Text-based language models leverage structured GUI metadata 
such as DOM trees and accessibility labels to generate symbolic commands, 
ranging from early page-centric agents \cite{nakano2021webgpt} to recent language-only planners 
that avoid raw pixel processing \cite{xu2023lemur}. 
Vision-centric agents incorporate screen imagery through two main strategies: 
grounding-focused methods that learn to associate natural-language references 
with bounding boxes or coordinate clicks \cite{gou2024navigating,wu2024atlas,xie2025scaling}, 
and end-to-end policies that directly translate screenshots into action sequences \cite{xu2024aguvis,qin2025ui,guo2025seed1}. 
Agent frameworks represent a third paradigm that enhances large language models 
with specialized components including vision encoders, 
hierarchical or search-based planners, episodic memory, 
and tool APIs to tackle long-horizon tasks 
requiring integrated perception, reasoning, and control \cite{agashe2024agent,agashe2025agent}.

\textbf{Reward Models for CUA. }
% 从oswolrd开始讲，就说这个benchmark是通过人工预先定义的脚本来验证agent执行任务的轨迹是否成功。
% 但是这样人工针对每个任务写脚本来验证轨迹是否成功，成本太高了，对于构造小规模的测试集尚且能够接受，
% 如果是要构造大规模的训练集，或者用于online RL，就捉襟见肘了。
% 于是，一些方法开始使用VLM作为验证轨迹的reward model。
%这些方法使用RM的方式可以分为两类。
% 一类是使用RM筛选成功的轨迹。
% seagent 训练了一个world state model来判断给定的轨迹是否成功以及agent首次犯错的步骤
% opencua利用Claude3.7逐步反思当前步骤，并且最终判断轨迹是否完成任务。
% gui-owl同时利用LLM和VLM生产two-channe step-level的critics，然后汇聚这些critics来判断轨迹是否成功。
% sea训练了一个step filteting model来过滤掉训练数据中中错误、冗余的轨迹。
% 另一类是使用RM为RL训练提供奖励信号。
% uitars-2在RL中使用uitars2本身作为ORM来验证通用web任务的
% seagent对world state model找到的first error step施加adversarial imitation punishment
OSWorld benchmark \cite{osworld} initially employed manually predefined scripts to verify agent trajectory success. 
However, writing custom verification scripts for each task incurs prohibitively high costs, 
making this approach inadequate for large-scale training datasets or online reinforcement learning systems. 
Consequently, recent approaches leverage Vision-Language Models (VLMs) 
as reward models for trajectory verification.
These reward model applications can be categorized into two primary paradigms. 
The first paradigm focuses on \textit{trajectory filtering}, 
where reward models identify successful trajectories. 
SEAgent~\cite{seagent} trains a world state model 
to determine trajectory success and identify the first error step. 
OpenCUA~\cite{opencua} employs Claude-3.7 for step-by-step reflection, 
ultimately judging whether trajectories accomplish their assigned tasks. 
GUI-OWL~\cite{guiowl} utilizes both LLMs and VLMs 
to generate two-channel step-level critics, 
aggregating these assessments to determine trajectory success. 
SEA~\cite{sea} develops a step filtering model 
to remove erroneous and redundant trajectories from training data. 
The second paradigm employs reward models 
to \textit{provide reward signals for reinforcement learning}. 
UITARS-2~\cite{uitars2} uses the UITARS-2 model itself 
as an outcome reward model (ORM) for general web task verification in RL settings. 
SEAgent \cite{seagent} applies adversarial imitation punishment 
to first error steps identified by their world state model.

\textbf{Reward Benchmarks.} 
There is some research evaluating reward models across multiple domains. 
RewardBench \cite{lambert2024rewardbench} established multi-domain evaluation for LLMs covering chat, reasoning, and safety. 
RM-Bench \cite{liu2024rmbench} introduced Best-of-N evaluation, 
while Multimodal RewardBench \cite{yasunaga2025multimodal} 
proposed evaluation frameworks for VLMs with expert-annotated triplets. 
For agents, AgentRewardBench \cite{lu2025agentrewardbench} 
evaluates Web agents but ignores desktop operations, 
while Agent-RewardBench \cite{men2025agent} covers multimodal agents 
but lacks CUA-specific capabilities like GUI positioning accuracy. 
In contrast, our CUARewardBench investigates reward model capabilities 
specifically for computer-using agents, 
covering desktop software operations and multi-step decision-making 
that previous benchmarks do not possess.

\subsection{Prompts Templates of Reward Models}
\label{sec:prompts}
% 对于ZeroGUI\cite{zerogui}和SE-WSM \cite{seagent}的prompt，我们照搬其开源代码的原始版本Figure \ref{}和Figure \ref{}。
% 对于OpenCUA reflector \cite{opencua_reflector}的prompt，
% 我们修改了其中与步骤CoT耦合的部分，使其能够适配于我们的实验环境，
% 原始的prompt和简化的prompt分别如Figure \ref{}和Figure \ref{}所示。
For ZeroGUI~\cite{zerogui} and SE-WSM~\cite{seagent} prompts, we directly adopt the original versions from their open-source implementations, as shown in Figure~\ref{fig:zerogui_prompt} and Figure~\ref{fig:sewsm_prompt}. 
For the OpenCUA reflector~\cite{opencua} prompt, we modified the components coupled with step-wise chain-of-thought reasoning to ensure compatibility with our experimental environment. The original and simplified prompts are presented in Figure~\ref{fig:enhanced_prm_prompt} and Figure~\ref{fig:step_evaluation_prompt}, respectively.

\begin{figure*}[htbp]
    \begin{tcolorbox}[colback=gray!10!white, colframe=gray!50!black, title={ORM Prompts of ZeroGUI},fontupper=\small]
    You are an expert at analyzing computer usage task completion from screenshots.
    
    You will be given a task instruction and a series of screenshots of the task execution. 
    Please analyze the screenshots and provide a detailed analysis of the task completion by following the steps below:
    
    1. First, analyze and understand the task instruction. Describe what should the screenshots look like if the task is completed successfully.
    
    2. Describe what you observe in each screenshot, analysis what actions were taken and what changes were made to the UI to achieve the task (or mistakes made).
    
    3. When you analyze the screenshots, please pay attention to the very detailed elements and changes in the UI. Every small detail may affect the final result.
    
    4. After all screenshots are analyzed, provide a overall reasoning about how the task was completed or failed at \textbf{the final state}. Make sure you have considered all demands of the task instruction.
    
    5. Determine if the task was completed at \textbf{the final state} (the last screenshot) successfully (score 1 for success, 0 for failure). If the task is completed during the process but not at the final state, it should be considered as failure (0 score).
    Provide your response strictly in the following format:
    
    TASK REQUIREMENT:
    
    [Your understanding of the task instruction]
    
    SCREENSHOT ANALYSIS:
    
    Screenshot 1:
    
    [Analysis of first screenshot]
    
    Screenshot 2:
    
    [Analysis of second screenshot]
    
    ...
    
    REASONING:
    
    [Your reasoning]
    
    FINAL ANSWER:
    
    [Your final answer]
    
    SCORE: [0/1]
    
    Now, please \textbf{strictly follow the format} and analyze the following screenshots (The last line should only be SCORE: [0/1], no other text):
    
    Task Instruction: \{instruction\}
    
    Screenshots (by order):
    \end{tcolorbox}
    \caption{ORM prompt of ZeroGUI \cite{zerogui}. The prompt instructs the vision-language model to analyze computer usage task completion through detailed screenshot examination and structured response formatting.}
    \label{fig:zerogui_prompt}
\end{figure*}

\begin{figure*}[htbp]
    \begin{tcolorbox}[colback=gray!10!white, colframe=gray!50!black, title={ORM and PRM Prompts of SE-WSM},fontupper=\small]
    You are an expert at analyzing computer usage task completion from screenshots. I am evaluating the performance of a UI agent. The images provided are \textbf{sequential keyframes} that represent the full execution trajectory of the agent when attempting to follow a command. These keyframes correspond to the instruction: \textbf{'\{instruction\}'}.
    
    Please thoroughly analyze the sequence to assess the following aspects:
    
    1. \textbf{Correctness} — Did the agent successfully complete the task as instructed?
    
    2. \textbf{Redundant Steps} — Identify any unnecessary or repeated actions that do not contribute to the goal.
    
    3. \textbf{Optimization} — Did the agent follow an efficient plan with a minimal number of steps?
    
    4. \textbf{First Error Step} — If the execution is incorrect or sub-optimal, determine the index of the \textbf{first keyframe where a mistake occurred}.
    
    5. \textbf{Error Analysis} — Provide a brief explanation of the mistake at that step.
    
    6. \textbf{Correct Action Suggestion} — Explain what the agent \textbf{should have done instead} at the point of error.
    
    \textbf{Important Instructions:}
    
    - The agent may have made progress toward the goal, but unless the task is \textbf{fully and correctly completed}, you must set 'Correctness' to \textbf{False}.
    
    - Be cautious in determining success. Missing confirmation screens, skipped inputs, or wrong UI elements clicked all count as errors.
    
    - Carefully examine all UI changes, button interactions, text entries, and any visual feedback in the screenshots.
    
    - Clearly indicate \textbf{which exact steps are redundant} (starting from 1).
    
    Once you finish the analysis, return your evaluation in the following dictionary format (include your step-by-step reasoning \textbf{above} the result):
    
    \textless analysis process\textgreater
    
    your step-by-step reasoning
    
    \textless /analysis process\textgreater
    
    \textless res\_dict\textgreater\\
    \{\\
    \quad ``Correctness'': True/False,\\
    \quad ``Redundant'': [step\_num, ...],\\
    \quad ``Optimized'': True/False,\\
    \quad ``First\_Error\_Step'': step\_num or None,\\
    \quad ``Error\_Type'': ``brief description of the mistake'',\\
    \quad ``Correct\_Action'': ``what should have been done instead''\\
    \}\\
    \textless /res\_dict\textgreater
    \end{tcolorbox}
    \caption{SE-WSM \cite{seagent} prompt template both for ORM and PRM evaluation. The prompt instructs the vision-language model to conduct multi-dimensional assessment including trajectory correctness, redundant steps identification, first error step detection, and correct action suggestions for both ORM and PRM evaluation.}
    \label{fig:sewsm_prompt}
\end{figure*}

\begin{figure*}[htbp]
    \begin{tcolorbox}[colback=gray!10!white, colframe=gray!50!black, title={PRM Prompts of OpenCUA Reflector},fontupper=\small]
    You are a judge of a computer-use agent. You will be given a task, the agent's history actions, agent last action and thought process with 2 screenshots.

    - Thought is the reasoning for the history steps and prediction for the next step.
    - Action is the summary of the code
    - Code is the code that will be executed.
    - The first screenshot is the observation of the last action and the second image is the computer state after executing the last action (code).

    \textbf{Task:} \{goal\}

    \textbf{History steps:} \{history\_steps\}

    \textbf{Last step:}

    \textbf{Thought:} \{thought\}

    \textbf{Action:} \{action\}

    \textbf{Code:} \{code\}

    Your response should include 3 parts:

    \textbf{1. Is the last step redundant:}
    - If the last step is doing unnecessary action or action that is not related to the task, for example, clicking irrelevant places, open irrelevant applications, or unnecessary scrolls, you should mark it as redundant.

    \textbf{2. Is the last step incorrect:}
    - If the action is related to the task but executing the code did not produce the expected change, you should mark it as incorrect.
    - If the action and the code do not align, you should mark it as incorrect. For example the action tries to click an element but failed according to the screenshot.
    - The last screenshot shows the application or window is not fully loaded, but the code is executed.
    - If there is any mistake in the thought action.

    \textbf{3. Reflection:}
    - You should first provide a natural summary of the visual changes between the last screenshot and the current screenshot. If there is no change, please mention it.
    - If the last step is correct and not redundant, you should then say the step is necessary and how it is effective.
    - If the last step is incorrect, you should then provide a clear explanation of the error.
    - If the last step is redundant, you should then provide a clear explanation.

    \textbf{YOUR RESPONSE MUST BE EXACTLY ONE VALID JSON OBJECT. NO MARKDOWN, NO EXTRA TEXT.}

    Here is the exact JSON structure you must follow:

    \textless res\_dict\textgreater\\
    \{\\
    \quad ``last\_step\_correct'': bool,\\
    \quad ``last\_step\_redundant'': bool,\\
    \quad ``reflection'': str\\
    \}\\
    \textless /res\_dict\textgreater
    \end{tcolorbox}
    \caption{original PRM Prompts of OpenCUA reflector \cite{opencua}. The prompt provides comprehensive step-level assessment including thought process analysis, action-code alignment verification, and structured JSON output format for systematic evaluation of agent decision-making processes.}
    \label{fig:enhanced_prm_prompt}
\end{figure*}

\begin{figure*}[htbp]
    \begin{tcolorbox}[colback=gray!10!white, colframe=gray!50!black, title={Simplified PRM Prompts of OpenCUA Reflector},fontupper=\small]
    You are a judge of a computer-use agent. Your role is to evaluate whether the agent's last action was redundant, incorrect, or appropriate for completing the given task. You will analyze screenshots showing the agent's history, the state before the last action, and the state after the last action, along with the raw code that was executed.

    You will be given a task, the agent's history screenshots, agent's last action code, and 2 screenshots showing before and after the last action.

    - The history screenshots show what happened before the last step, helping you understand the agent's previous progress and context.
    - The last action is represented by the raw code that was executed.
    - The before and after screenshots show the state immediately before and after executing the last action code.
    - Note: If there is mouse related code that needs coordinates, the center of the red circle in the before screenshot shows the position. But do not mention the red circle or red dot in any part of your response.
    - You will be provided with: history screenshots (showing previous steps), one screenshot before executing the last action, and one screenshot after executing the last action.
    - You should focus on the differences between the before and after screenshots to understand what the last action accomplished, while using history screenshots to understand the context and detect redundancy.

    \textbf{Task:} \{instruction\}

    \textbf{History screenshots:} \textless image\textgreater

    \textbf{Screenshots before and after the last action:}  \textless image\textgreater

    \textbf{Last action code:} Step \{step\_index\}: \{action\_code\}

    Your response should include 3 parts:

    \textbf{1. Is the last step redundant:}
    - If the last step is doing unnecessary action or action that is not related to the task, for example, clicking irrelevant places, open irrelevant applications, or unnecessary scrolls, you should mark it as redundant.
    - If the last step is a repeat of a former step based on the history screenshots, you should mark it as redundant.
    - Too many scrolls or drags of the scroll bar, or too many clicks of the same button, or too many clicks of the same element, you should mark it as redundant.

    \textbf{2. Is the last step incorrect:}
    - If the action is related to the task but executing the code did not produce the expected change, you should mark it as incorrect.
    - If the code execution failed or did not work as intended based on the before/after screenshots, you should mark it as incorrect.
    - The after screenshot shows the application or window is not fully loaded, but the code was executed.
    - If there is any clear mistake in the action based on what the code was trying to accomplish.
    - You should carefully examine the click/drag related actions. In many cases, the code wants to click a target, but it doesn't match the element at the center of the red circle in the before screenshot.

    \textbf{3. Reflection:}
    - You should first provide a natural summary of the visual changes between the before and after screenshots. If there is no change, please mention it.
    - If the last step is correct and not redundant, you should then say the step is necessary and how it is effective toward completing the task.
    - If the last step is incorrect, you should then provide a clear explanation of the error.
    - If the last step is redundant, you should then provide a clear explanation of why it's unnecessary given the history.

    Once you finish the analysis, return your evaluation in the following dictionary format (include your step-by-step Reflection \textbf{above} the result):

    \textless analysis process\textgreater

    [your step-by-step reflection]

    \textless /analysis process\textgreater

    \textless res\_dict\textgreater\\
    \{\\
    \quad ``last\_step\_correct'': bool,\\
    \quad ``last\_step\_redundant'': bool,\\
    \quad ``reflection'': str\\
    \}\\
    \textless /res\_dict\textgreater
    \end{tcolorbox}
    \caption{Simplified PRM Prompts of OpenCUA reflector \cite{opencua}. 
    The prompt instructs the vision-language model 
    to assess individual agent actions by analyzing visual changes 
    between consecutive screenshots 
    and determining whether the latest action is correct, redundant, or necessary for task completion.}
    \label{fig:step_evaluation_prompt}
\end{figure*}

\subsection{Supplementary Results for Additional Task Categories}

This section presents the experimental results 
for the remaining 5 task categories 
that were not included in the main paper due to space constraints. 
These results complement the overall performance 
and 5 selected categories (VS Code, GIMP, LibreOffice Writer, Chrome, 
and Multi-apps) shown in Tables~\ref{tab:reward_model_performance} 
and \ref{tab:action_reward_model_performance}, 
providing complete coverage of all 10 software categories in CUARewardBench.

The supplementary categories 
include LibreOffice Calc, LibreOffice Impress, VLC, Thunderbird, 
and OS operations. T
hese results maintain consistency with the patterns 
observed in the main text, 
further validating our key findings.

\begin{table*}[t]
    \centering
    \scriptsize  % 改为更小的字体
    \setlength{\tabcolsep}{3.8pt}  % 减小列间距以容纳更多列
    \begin{tabular}{lcccccc|cc|cc|cc|cc|cc}
    \toprule
    \multirow{2}{*}{Reward Model} & \multicolumn{6}{c|}{Overall} & \multicolumn{2}{c|}{vlc} & \multicolumn{2}{c|}{os} & \multicolumn{2}{c|}{thunderbird} & \multicolumn{2}{c|}{impress} & \multicolumn{2}{c}{calc} \\
    & P & NPV & R & S & OA & & P & NPV & P & NPV & P & NPV & P & NPV & P & NPV \\
    \midrule
    \multicolumn{17}{c}{\textit{zerogui}} \\
    \midrule
    Qwen2.5VL-7B & 60.8 & 65.0 & 74.8 & 49.2 & 62.1 & & 71.4 & 66.7 & 61.1 & 62.5 & 50.0 & 50.0 & 47.6 & 42.9 & 65.0 & 71.4 \\
    Qwen2.5VL-32B & 70.9 & 76.1 & 80.6 & 65.2 & 72.8 & & 72.7 & 55.6 & 64.3 & 58.3 & \textbf{100.0} & \textbf{80.0} & 66.7 & 77.8 & 68.2 & 83.3 \\
    Qwen2.5VL-72B & 70.0 & 72.1 & 75.5 & 66.2 & 71.0 & & 83.3 & 50.0 & 66.7 & 63.6 & \textbf{100.0} & 72.7 & 63.2 & 77.8 & 63.2 & 66.7 \\
    GLM-4.5V-106B & \textbf{76.8} & \textbf{90.4} & \textbf{92.8} & \textbf{70.7} & \textbf{82.0} & & 91.7 & \textbf{87.5} & 68.4 & \textbf{85.7} & \textbf{100.0} & \textbf{80.0} & \textbf{70.6} & \textbf{81.8} & \textbf{78.9} & \textbf{86.7} \\
    GUI-OWL-7B & 69.0 & 68.8 & 71.9 & 65.6 & 68.4 & & 75.0 & 50.0 & \textbf{71.4} & 66.7 & 80.0 & 63.6 & 58.3 & 56.2 & 62.5 & 61.1 \\
    GUI-OWL-32B & 69.0 & 73.7 & 78.4 & 63.2 & 71.0 & & \textbf{100.0} & 61.5 & 68.8 & 70.0 & 75.0 & 75.0 & 50.0 & 50.0 & 59.1 & 66.7 \\
    \midrule
    \multicolumn{17}{c}{\textit{sewsm}} \\
    \midrule
    Qwen2.5VL-7B & 63.1 & 57.1 & 50.4 & 69.2 & 59.6 & & 50.0 & 35.7 & 75.0 & 55.6 & 60.0 & 54.5 & 66.7 & 62.5 & 55.6 & 56.2 \\
    Qwen2.5VL-32B & 80.0 & 71.7 & 69.1 & 82.0 & 75.4 & & 75.0 & 50.0 & 81.8 & 66.7 & \textbf{100.0} & 72.7 & \textbf{85.7} & \textbf{85.7} & 78.6 & 70.0 \\
    Qwen2.5VL-72B & 78.6 & 74.5 & 74.1 & 78.9 & 76.5 & & 77.8 & 54.5 & 73.3 & 72.7 & 66.7 & 60.0 & 75.0 & 68.8 & 82.4 & \textbf{82.4} \\
    GLM-4.5V-106B & \textbf{82.9} & \textbf{77.6} & \textbf{77.0} & 83.5 & \textbf{80.1} & & \textbf{100.0} & 53.3 & \textbf{86.7} & \textbf{90.9} & \textbf{100.0} & 72.7 & 75.0 & 83.3 & \textbf{91.7} & 72.7 \\
    GUI-OWL-7B & 68.4 & 72.4 & 76.8 & 63.2 & 69.9 & & 87.5 & \textbf{58.3} & 64.7 & 75.0 & \textbf{100.0} & \textbf{80.0} & 50.0 & 50.0 & 63.6 & 75.0 \\
    GUI-OWL-32B & 75.0 & 71.4 & 71.2 & 75.2 & 73.2 & & 85.7 & 53.8 & 73.3 & 72.7 & 66.7 & 71.4 & 62.5 & 66.7 & 72.7 & 60.9 \\
    SE-WSM-7B & 70.0 & 52.2 & 20.1 & \textbf{91.0} & 54.8 & & \textbf{100.0} & 57.1 & 33.3 & 40.0 & 0.0 & 50.0 & 0.0 & 48.1 & 33.3 & 48.4 \\
    \midrule
    \multicolumn{17}{c}{\textit{voting-majority}} \\
    \midrule
    G106-s 2runs & 84.3 & 70.7 & 65.5 & 87.2 & 76.1 & & \textbf{100.0} & 53.3 & 83.3 & 71.4 & \textbf{100.0} & 66.7 & 83.3 & 75.0 & 90.0 & 66.7 \\
    Q32-s + G106-s & \textbf{90.1} & 68.5 & 59.0 & \textbf{93.2} & 75.7 & & \textbf{100.0} & 50.0 & \textbf{90.0} & 68.8 & \textbf{100.0} & 66.7 & \textbf{100.0} & \textbf{87.5} & 87.5 & 61.5 \\
    \makecell[l]{Q32-s + G106-s \\+ G106-z} & 81.6 & \textbf{84.8} & \textbf{86.3} & 79.7 & \textbf{83.1} & & \textbf{100.0} & \textbf{61.5} & 85.7 & \textbf{83.3} & \textbf{100.0} & \textbf{80.0} & 75.0 & 83.3 & \textbf{93.8} & \textbf{88.9} \\
    \midrule
    \multicolumn{17}{c}{\textit{voting-strict\_unanimous}} \\
    \midrule
    G106-s 2runs & 84.3 & 82.3 & \textbf{65.5} & \textbf{76.7} & \textbf{71.0} & & \textbf{100.0} & 53.8 & \textbf{90.0} & 90.0 & \textbf{100.0} & \textbf{80.0} & 83.3 & 85.7 & \textbf{90.0} & 82.4 \\
    Q32-s + G106-s & \textbf{90.1} & 84.2 & 59.0 & 72.2 & 65.4 & & \textbf{100.0} & 54.5 & \textbf{90.0} & 90.0 & \textbf{100.0} & \textbf{80.0} & \textbf{100.0} & 80.0 & 87.5 & 87.5 \\
    \makecell[l]{Q32-s + G106-s \\+ G106-z} & 89.8 & \textbf{93.3} & 56.8 & 63.2 & 59.9 & & \textbf{100.0} & \textbf{83.3} & \textbf{90.0} & \textbf{100.0} & \textbf{100.0} & \textbf{80.0} & \textbf{100.0} & \textbf{87.5} & 85.7 & \textbf{91.7} \\
    \bottomrule
    \end{tabular}
    \caption{Supplementary results for Table \ref{tab:reward_model_performance}, showing performance of outcome reward models (ORM) on the remaining task categories (vlc, os, thunderbird, impress, calc). 
    Results show precision (P) and negative predictive value (NPV) for trajectory success evaluation 
    under multiple prompt configurations: \textit{zerogui}, \textit{sewsm}, \textit{voting-majority}, and \textit{voting-strict\_unanimous}. 
    The Overall metrics remain the same as in Table \ref{tab:reward_model_performance}.} 
    \label{tab:appendix_reward_model_performance}
\end{table*}

\begin{table*}[t]
    \centering
    \scriptsize  % 改为更小的字体
    \setlength{\tabcolsep}{3.8pt}  % 减小列间距
    \begin{tabular}{lccccc|cc|cc|cc|cc|cc}
    \toprule
    \multirow{2}{*}{Reward Model} & \multicolumn{5}{c|}{Overall} & \multicolumn{2}{c|}{vlc} & \multicolumn{2}{c|}{os} & \multicolumn{2}{c|}{thunderbird} & \multicolumn{2}{c|}{impress} & \multicolumn{2}{c}{calc} \\
    & P & NPV & R & S & OA & P & NPV & P & NPV & P & NPV & P & NPV & P & NPV \\
    \midrule
    \multicolumn{16}{c}{\textit{opencua\_reflector}} \\
    \midrule
    Qwen2.5VL-7B & 54.4 & 49.4 & 53.8 & 50.0 & 52.0 & 62.5 & 40.0 & 47.1 & 54.5 & 30.0 & 44.4 & 71.4 & 54.5 & 55.6 & 47.4 \\
    Qwen2.5VL-32B & 60.3 & 64.8 & 79.8 & 41.5 & 61.7 & 68.4 & 57.1 & 55.0 & 66.7 & 42.9 & 60.0 & 71.4 & \textbf{66.7} & 50.0 & 38.5 \\
    Qwen2.5VL-72B & 58.5 & 64.8 & 83.0 & 34.8 & 60.1 & 75.0 & 83.3 & 45.5 & 50.0 & 41.7 & 57.1 & 60.0 & 50.0 & 58.8 & 62.5 \\
    GLM-4.5V-106B & 64.0 & \textbf{78.5} & \textbf{89.0} & 44.5 & \textbf{67.9} & 72.7 & \textbf{100.0} & 47.8 & 60.0 & 50.0 & \textbf{80.0} & 72.2 & 61.1 & \textbf{64.5} & \textbf{100.0} \\
    GUI-OWL-7B & \textbf{64.6} & 61.8 & 67.0 & \textbf{59.1} & 63.3 & \textbf{76.5} & 66.7 & \textbf{55.6} & 70.0 & \textbf{71.4} & 75.0 & \textbf{76.5} & 63.2 & 61.9 & 56.2 \\
    GUI-OWL-32B & 61.6 & 61.5 & 71.4 & 50.6 & 61.6 & 70.0 & 66.7 & 55.0 & \textbf{75.0} & 42.9 & 58.3 & 61.9 & 53.3 & 57.9 & 50.0 \\
    \midrule
    \multicolumn{16}{c}{\textit{sewsm}} \\
    \midrule
    Qwen2.5VL-7B & 56.7 & 54.2 & 67.0 & 43.3 & 55.8 & 73.7 & 71.4 & 42.1 & 44.4 & 33.3 & 42.9 & \textbf{75.0} & \textbf{68.8} & 51.7 & 37.5 \\
    Qwen2.5VL-32B & 65.3 & 62.4 & 67.8 & 59.8 & 64.0 & \textbf{76.5} & 66.7 & 53.3 & 57.1 & 55.6 & 70.0 & 58.3 & 50.0 & 47.8 & 35.7 \\
    Qwen2.5VL-72B & 58.7 & 67.1 & 85.2 & 33.5 & 60.7 & 72.7 & \textbf{100.0} & 45.0 & 50.0 & 50.0 & 80.0 & 60.0 & 54.5 & 54.8 & 50.0 \\
    GLM-4.5V-106B & \textbf{69.5} & 64.2 & 66.1 & \textbf{67.7} & \textbf{66.9} & \textbf{76.5} & 66.7 & \textbf{53.8} & 56.2 & 57.1 & 66.7 & 61.9 & 53.3 & \textbf{73.3} & \textbf{59.1} \\
    GUI-OWL-7B & 56.6 & 64.2 & 86.8 & 26.2 & 58.1 & 66.7 & 60.0 & 48.0 & \textbf{66.7} & \textbf{70.0} & \textbf{88.9} & 58.6 & 57.1 & 56.7 & 57.1 \\
    GUI-OWL-32B & 57.4 & \textbf{68.8} & \textbf{89.0} & 26.8 & 59.5 & 66.7 & 60.0 & 50.0 & \textbf{66.7} & 46.7 & 75.0 & 57.6 & 66.7 & 53.8 & 45.5 \\
    SE-WSM-7B & 58.7 & 52.0 & 48.4 & 62.2 & 54.9 & 75.0 & 60.0 & 30.8 & 40.0 & 33.3 & 50.0 & 56.2 & 45.0 & 50.0 & 43.5 \\
    \midrule
    \multicolumn{16}{c}{\textit{voting-majority}} \\
    \midrule
    G106-s 2runs & 73.6 & 62.8 & 59.6 & 76.2 & 67.4 & \textbf{86.7} & 72.7 & \textbf{60.0} & \textbf{57.9} & \textbf{66.7} & 62.5 & 61.9 & 53.3 & \textbf{76.9} & 58.3 \\
    Q32-s + G106-s & \textbf{75.6} & 60.5 & 52.5 & \textbf{81.1} & 66.0 & 85.7 & 66.7 & 55.6 & 55.0 & 57.1 & \textbf{66.7} & 58.8 & 47.4 & 66.7 & 50.0 \\
    \makecell[l]{Q32-s + G106-s \\+ G106-o} & 68.4 & \textbf{71.5} & \textbf{78.6} & 59.8 & \textbf{69.7} & 77.8 & \textbf{75.0} & 47.1 & 54.5 & 50.0 & 63.6 & \textbf{62.5} & \textbf{58.3} & 59.3 & \textbf{60.0} \\
    \midrule
    \multicolumn{16}{c}{\textit{voting-strict\_unanimous}} \\
    \midrule
    G106-s 2runs & 73.6 & 64.2 & \textbf{59.6} & \textbf{57.9} & \textbf{58.8} & 86.7 & 57.1 & \textbf{60.0} & 50.0 & \textbf{66.7} & 63.6 & 61.9 & 54.5 & \textbf{76.9} & 64.7 \\
    Q32-s + G106-s & 75.6 & 69.1 & 52.5 & 46.3 & 49.6 & 85.7 & 66.7 & 55.6 & 60.0 & 57.1 & 70.0 & 58.8 & \textbf{62.5} & 66.7 & 50.0 \\
    \makecell[l]{Q32-s + G106-s \\+ G106-o} & \textbf{81.7} & \textbf{85.1} & 48.9 & 24.4 & 37.3 & \textbf{92.3} & \textbf{100.0} & 55.6 & \textbf{100.0} & 57.1 & \textbf{100.0} & \textbf{80.0} & 57.1 & 75.0 & \textbf{100.0} \\
    \bottomrule
    \end{tabular}
    \caption{Supplementary results for Table \ref{tab:action_reward_model_performance}, showing performance of process reward models (PRM) on the remaining task categories (vlc, os, thunderbird, impress, calc). 
    Results show precision (P) and negative predictive value (NPV) for step-level correctness assessment 
    under multiple prompt configurations: \textit{opencua\_reflector}, \textit{sewsm}, \textit{voting-majority}, and \textit{voting-strict\_unanimous}. 
    The Overall metrics remain the same as in Table \ref{tab:action_reward_model_performance}.}
    \label{tab:appendix_action_reward_model_performance}
\end{table*}

\subsection{Use of Large Language Models}
\label{sec:llm_usage}

In accordance with ICLR 2026 policy, we disclose that Large Language Models (LLMs) were used in the preparation of this paper. 
Specifically, LLMs were employed to  polish the writing, including improving clarity, grammar, and overall presentation of the content. 
The LLMs were used solely as writing assistance tools and did not contribute to the conceptual or technical aspects of the research.

\end{document}